\pdfoutput=1
\documentclass[iop, apj, twocolappendix, numberedappendix]{emulateapj}

\newif\iflocal
\localfalse

\usepackage{xspace}
\usepackage{amsmath}
\usepackage{framed} 
\usepackage{txfonts}
\usepackage{epstopdf}
\usepackage{color}
\usepackage{rotating}
\usepackage{natbib}
\usepackage{ulem}
\usepackage{xspace}
\usepackage[colorlinks=true,urlcolor=black,linkcolor=blue,citecolor=blue]{hyperref}
\usepackage{sistyle}
\SIthousandsep{,}

\iflocal
\def\includedir{/Users/benedito/University/docs/latex}
\def\figdir{figs}
\else
\def\includedir{.}
\def\figdir{.}
\fi

\usepackage{etoolbox}
\makeatletter
\patchcmd{\NAT@citex}
  {\@citea\NAT@hyper@{\NAT@nmfmt{\NAT@nm}\NAT@date}}
  {\@citea\NAT@nmfmt{\NAT@nm}\NAT@hyper@{\NAT@date}}
  {}
  {}
\patchcmd{\NAT@citex}
  {\@citea\NAT@hyper@{%
     \NAT@nmfmt{\NAT@nm}%
     \hyper@natlinkbreak{\NAT@aysep\NAT@spacechar}{\@citeb\@extra@b@citeb}%
     \NAT@date}}
  {\@citea\NAT@nmfmt{\NAT@nm}%
   \NAT@aysep\NAT@spacechar%
   \NAT@hyper@{\NAT@date}}
  {}
  {}
\patchcmd{\NAT@citex}
  {\@citea\NAT@hyper@{%
     \NAT@nmfmt{\NAT@nm}%
     \hyper@natlinkbreak{\NAT@spacechar\NAT@@open\if*#1*\else#1\NAT@spacechar\fi}%
       {\@citeb\@extra@b@citeb}%
     \NAT@date}}
  {\@citea\NAT@nmfmt{\NAT@nm}%
   \NAT@spacechar\NAT@@open\if*#1*\else#1\NAT@spacechar\fi%
   \NAT@hyper@{\NAT@date}}
  {}
  {}
\makeatother

\def\gtsima{$\; \buildrel > \over \sim \;$}
\def\ltsima{$\; \buildrel < \over \sim \;$}
\def\prosima{$\; \buildrel \propto \over \sim \;$}
\def\gsim{\lower.7ex\hbox{\gtsima}}
\def\lsim{\lower.7ex\hbox{\ltsima}}
\def\simgt{\lower.7ex\hbox{\gtsima}}
\def\simlt{\lower.7ex\hbox{\ltsima}}
\def\simpr{\lower.7ex\hbox{\prosima}}

\newcommand{\kpch}{{h^{-1}{\rm kpc}}}
\newcommand{\mpch}{h^{-1}{\rm {Mpc}}}

\newcommand{\msun}{{M_{\odot}}}
\newcommand{\msunh}{h^{-1} M_\odot}

\def\LCDM{$\Lambda$CDM\xspace}

\def\sparta{\textsc{Sparta}\xspace}
\def\moria{\textsc{Moria}\xspace}

\def\colossus{\textsc{Colossus}\xspace}
\def\rockstar{\textsc{Rockstar}\xspace}
\def\consistenttrees{\textsc{Consistent-Trees}\xspace}

\def\planck{Planck\xspace}
\def\wmap{WMAP7\xspace}

\def\erebos{Erebos\xspace}

\def\deltac{\delta_{\rm c}}

\def\vmax{V_{\rm max}}

\def\fsub{f_{\rm sub}}
\def\ffb{f_{\rm flyby}}

\def\neff{n_{\rm eff}}

\def\rvir{R_{\rm vir}}

\def\mtom{M_{\rm 200m}}
\def\rtom{R_{\rm 200m}}

\def\nutom{\nu_{\rm 200m}}

\def\mtoc{M_{\rm 200c}}
\def\rtoc{R_{\rm 200c}}

\def\mfoc{M_{\rm 500c}}
\def\rfoc{R_{\rm 500c}}

\def\rsp{R_{\rm sp}}

\def\msp{M_{\rm sp}}

\shorttitle{Flybys, Orbits, Splashback}
\shortauthors{Flybys, Orbits, Splashback}

\journalinfo{The Astrophysical Journal, {\rm 909:112 (16pp), 2021 March 10}}
\submitted{Received 2020 July 18; revised 2020 December 23; accepted 2021 January 5; published 2021 March 10}

\begin{document}


\iflocal
\def\figdir{figs}
\else
\def\figdir{.}
\fi


\defcitealias{diemer_13_scalingrel}{DKM13}
\defcitealias{diemer_14}{DK14}
\defcitealias{diemer_15}{DK15}
\defcitealias{diemer_17_sparta}{Paper I}
\defcitealias{diemer_17_rsp}{Paper II}
\defcitealias{diemer_20_catalogs}{Paper III}


\title{Flybys, orbits, splashback: subhalos and the importance of the halo boundary}
\author{Benedikt Diemer$^{1,2}$}
\altaffiliation{$^2$NHFP Einstein Fellow}
\affil{
$^1$Department of Astronomy, University of Maryland, College Park, MD 20742, USA; \href{mailto:diemer@umd.edu}{diemer@umd.edu}
}


\begin{abstract}
The classification of dark matter halos as isolated hosts or subhalos is critical for our understanding of structure formation and the galaxy--halo connection. Most commonly, subhalos are defined to reside inside a spherical overdensity boundary such as the virial radius. The resulting host--subhalo relations depend sensitively on the somewhat arbitrary overdensity threshold, but the impact of this dependence is rarely quantified. The recently proposed splashback radius tends to be larger and to include more subhalos than even the largest spherical overdensity boundaries. We systematically investigate the dependence of the subhalo fraction on the radius definition and show that it can vary by factors of unity between different spherical overdensity definitions. Using splashback radii can yet double the abundance of subhalos compared to the virial definition. We also quantify the abundance of flyby (or backsplash) halos, hosts that used to be subhalos in the past. We show that the majority of these objects are mislabeled satellites that are naturally classified as subhalos when we use the splashback radius. We show that the subhalo fraction can be understood as a universal function of only peak height and the slope of the linear power spectrum. We provide a simple fitting function that captures our simulation results to 20\% accuracy across a wide range of halo masses, redshifts, and cosmologies. Finally, we demonstrate that splashback radii significantly change our understanding of satellite and flyby galaxies in the Local Group.
\end{abstract}



\section{Introduction}
\label{sec:intro}

In the \LCDM picture of structure formation, dark matter collapses into halos under its own gravity, pulling in baryons that form a galaxy at the center of the halo \citep{rees_77}. Even before the inception of this theory, galaxies were thought to collide with each other \citep[e.g.][]{toomre_72, ostriker_75}. These mergers are explained by the idea of hierarchical structure formation, where small halos form first, fall into a larger host halo and become subhalos \citep[e.g.][]{white_78, bond_91, lacey_93}. Given that structure is roughly self-similar across size scales, the large number of satellites observed in galaxy clusters implies that even galactic halos must contain an abundance of substructure \citep{katz_93, moore_99_subs, klypin_99_missing, kravtsov_04_satellites}. Even though subhalos are eventually disrupted and merge with their host, they can live for a substantial fraction of a Hubble time because of the long dynamical timescale of halos.

The existence of subhalos complicates the picture of structure formation significantly, both observationally and theoretically. Satellite galaxies live in physically different environments with higher density and experience disruptive processes such as ram pressure stripping \citep{abadi_99}. Similarly, their subhalos tend to be tidally disrupted, causing a sharp decline in their mass. For many purposes, it is thus important to distinguish between host and subhalos. Theoretically, for instance, the distinction matters when computing mass functions (of host halos) or assembly bias \citep{villareal_17}. Observationally, it matters when estimating cluster masses based on their richness \citep[the number of satellites, e.g.,][]{rozo_10, des_20_clusters} or for effects such as galactic conformity \citep{kauffmann_13}. The connection between observable galaxies and the dark matter universe is often established through prescriptions such as halo occupation distributions, subhalo abundance matching, or semianalytic models, most of which rely on a host--subhalo classification to some extent \citep[see][for a review]{wechsler_18}.

\newcommand{\figsizeviz}{0.505}
\begin{figure*}
\centering
\includegraphics[trim =  0mm 0mm 0mm 0mm, clip, scale=\figsizeviz]{\figdir/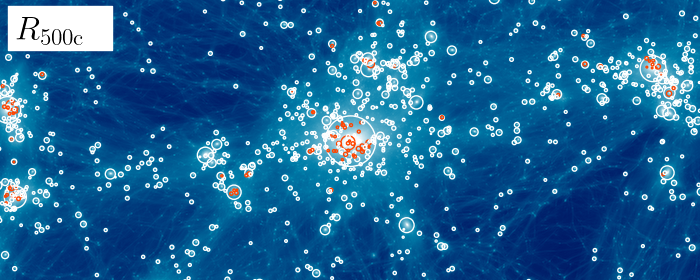}
\vspace{0.4mm}
\includegraphics[trim =  0mm 0mm 0mm 0mm, clip, scale=\figsizeviz]{\figdir/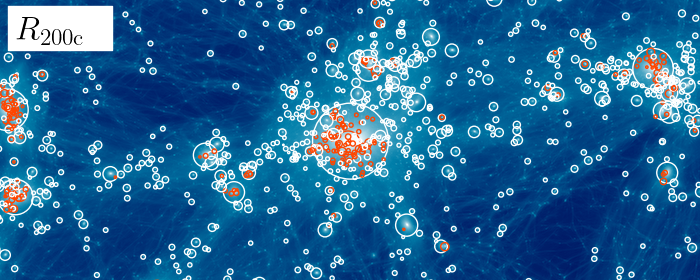}
\includegraphics[trim =  0mm 0mm 0mm 0mm, clip, scale=\figsizeviz]{\figdir/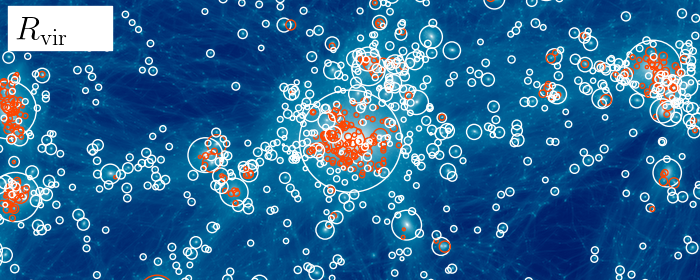}
\vspace{0.4mm}
\includegraphics[trim =  0mm 0mm 0mm 0mm, clip, scale=\figsizeviz]{\figdir/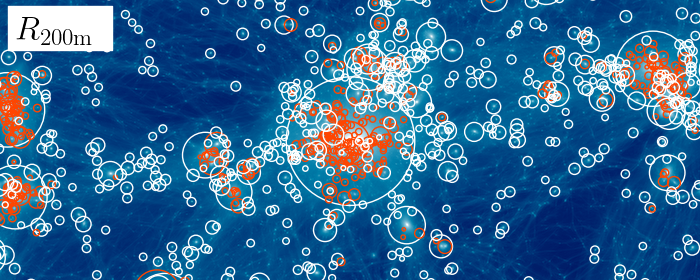}
\includegraphics[trim =  0mm 0mm 0mm 0mm, clip, scale=\figsizeviz]{\figdir/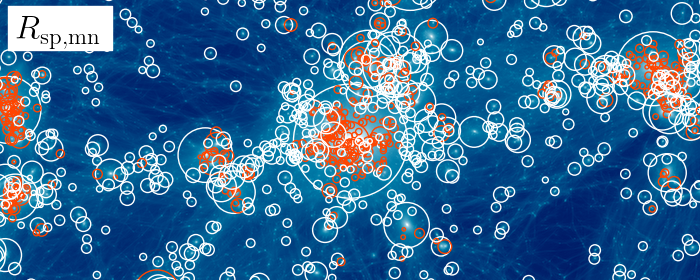}
\vspace{0.4mm}
\includegraphics[trim =  0mm 0mm 0mm 0mm, clip, scale=\figsizeviz]{\figdir/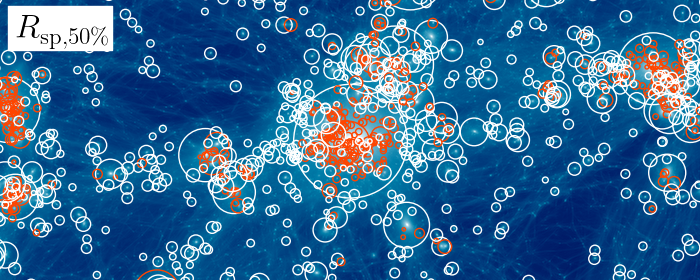}
\includegraphics[trim =  0mm 0mm 0mm 0mm, clip, scale=\figsizeviz]{\figdir/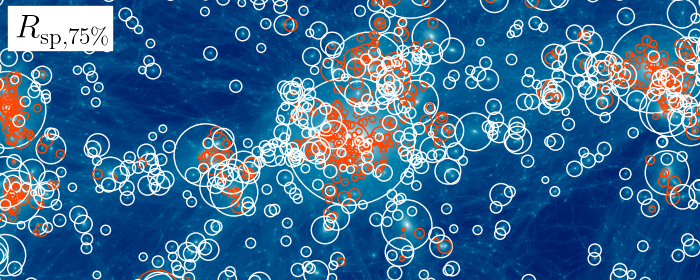}
\includegraphics[trim =  0mm 0mm 0mm 0mm, clip, scale=\figsizeviz]{\figdir/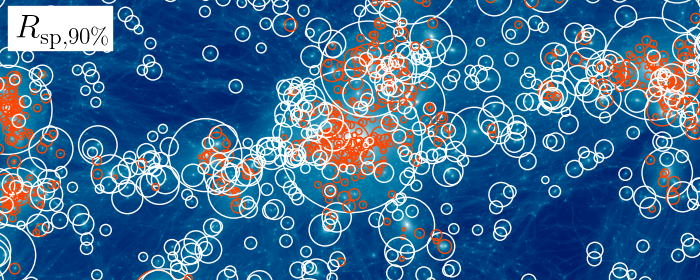}
\caption{Impact of the halo radius definition on the abundance of subhalos. Each panel shows the same background image, the projected density in a slab around a massive halo in the L0125-WMAP7 simulation. Circles show the radii of all hosts (white) and subhalos (orange) that reached $N_{\rm 200m} \geq 500$ at any point in their history. The first four panels show bound-only $\rfoc$, $\rtoc$, $\rvir$, and $\rtom$ as computed by \rockstar. The second four panels show the splashback radii corresponding to the mean, median, 75\%, and 90\% of the particle apocenter distribution. The splashback radius is not defined for subhalos; we replace it with $R_{\rm 200m,bnd}$. The visualization of the density field was created using the gotetra code by P. Mansfield (\href{https://github.com/phil-mansfield/gotetra}{https://github.com/phil-mansfield/gotetra}), which uses a tetrahedron-based estimate of the density field \citep{abel_12, kaehler_12, hahn_13}.}
\label{fig:viz}
\end{figure*}

To define a subhalo's membership in a larger host, we could, for instance, include all subgroups in the halo's friends-of-friends (FOF) group \citep{davis_85, springel_01_subfind}. However, this definition depends on an arbitrary linking length and is hard to infer from observations \citep{more_11_fof}. Instead, the most common definition of a subhalo is that it lives inside the spherical overdensity (SO) radius of a larger halo. These radii enclose an overdensity that is set to a fixed or varying multiple of the critical or matter density of the universe, leading to definitions such as $\rtoc$, $\rvir$, or $\rtom$ \citep[e.g.,][]{lacey_94}. SO definitions are easy to compute, but differences in the chosen overdensity lead to significant differences in radius and thus in the number of subhalos (see Figure~\ref{fig:viz} for a visualization). Although understood by practitioners, this difference is rarely quantified because halo catalogs typically give subhalo relations according to only one definition.

Moreover, there are reasons to question whether commonly used SO definitions truly capture the physical nature of halos. For instance, infalling subhalos and galaxies begin to lose mass at about two host virial radii on average, indicating that the sphere of influence of host halos is larger than suggested by $\rvir$ \citep{bahe_13, behroozi_14}. Another indication that SO radii do not include the full extent of halos is provided by a large population of ``isolated'' halos that used to be subhalos but now orbit outside their former host's virial radius. Often labeled ``backsplash halos'' or ``ejected satellites,'' the vast majority of these systems will eventually fall back onto their past host \citep{balogh_00, mamon_04, gill_05, pimbblet_11, wetzel_14_ejected, buck_19, haggar_20, knebe_20}, although there is also a (much smaller) population of genuine, temporary ``flyby'' events and interactions \citep{sales_07, ludlow_09, sinha_12, lhuillier_17, an_19_flyby}. To avoid confusion, we summarily refer to all of these phenomena as ``flyby.'' We argue that most flyby halos are misclassified subhalos, a distinction that matters because flyby halos and their galaxies may carry significant imprints of their interaction with the larger host \citep{knebe_11_backsplash, muriel_14}. Moreover, flyby halos can lead to a double-counting of mergers \citep{xie_15, benson_17} and cause spurious assembly bias signals \citep{sunayama_16, villareal_17, mansfield_20_ab, tucci_20}. 

To mitigate these issues, the splashback\footnote{The nomenclature can be misleading. Flyby halos are often referred to as ``backsplash halos,'' which is the original root of the term ``splashback radius.'' However, while the former refers to subhalos that are close to the apocenter of their orbit and thus outside of the virial radius (or a similarly defined boundary), the splashback radius refers to dark matter particles as well as subhalos and aims to include all orbits by construction. We avoid the term ``backsplash halos'' in this paper to avoid confusion.} radius, $\rsp$, has been put forward as a physically motivated definition of the halo boundary \citep{diemer_14, adhikari_14, more_15}. By definition, $\rsp$ corresponds to the apocenter of particles on their first orbit, which, in spherical symmetry, would include the orbits of all particles and subhalos and separate infalling from orbiting material \citep{fillmore_84, bertschinger_85, adhikari_14, shi_16_rsp}. In practice, measuring $\rsp$ is more difficult due to non-sphericity and interactions between halos, but it can be detected based on its accompanying sharp drop in the density field \citep{mansfield_17} or from particle dynamics \citep[][hereafter \citetalias{diemer_17_sparta}]{diemer_17_sparta}. Observationally, the splashback radius has been measured as a drop in the galaxy density and weak lensing signal around clusters \citep[e.g.,][]{more_16, chang_18}. \citet{baxter_17} demonstrated that the drop in density is predominantly caused by red cluster galaxies, while the infalling, bluer population follows a smooth profile, supporting the notion that the splashback radius separates infalling galaxies from those that have orbited at least once \citep[][see Figure~\ref{fig:trees} for a visualization]{aung_20_phasespace, bakels_20, tomooka_20}. While the relationship between SO and splashback radii was investigated in \citet{more_15} and \citet[][hereafter \citetalias{diemer_17_rsp}]{diemer_17_rsp}, the impact of $\rsp$ on the subhalo assignment has yet to be quantified systematically (although \citealt{mansfield_20_ab} found it to be substantial).

In this paper, we quantify the abundance of subhalos and flyby halos as a function of the halo boundary definition, considering a number of SO and splashback definitions. We show that subhalo fractions can be understood statistically as a function of the initial power spectrum and present a universal fitting formula built on this insight. Our results are based on the halo catalogs and merger trees presented by \citet[][hereafter \citetalias{diemer_20_catalogs}]{diemer_20_catalogs}, which are publicly available at \href{http://www.benediktdiemer.com/data}{benediktdiemer.com/data}. Throughout the paper, we follow the notation of \citetalias{diemer_20_catalogs}.

The paper is structured as follows. In Section~\ref{sec:method}, we describe our simulations and halo catalogs, deferring numerical details to the Appendix. We present our results in Section~\ref{sec:results} and discuss their possible applications in Section~\ref{sec:discussion}. We summarize our conclusions in Section~\ref{sec:conclusion}.  

In order to keep the paper focused on the effects of the radius definition, we sidestep a number of important issues. First, we ignore the unphysical numerical disruption of subhalos in $N$-body simulations, also known as the ``overmerging problem''  \citep[e.g.,][]{carlberg_94, vankampen_95, moore_96, moore_99_collapse, klypin_99_overmerging}. This issue still affects modern simulations, including those used in this work \citep{vandenbosch_18_subs1}. Second, we do not tackle the question of how to define the mass of a subhalo; instead, we give results for a variety of definitions such as the bound-only mass, peak mass, and circular velocity. We return to this issue in Diemer \& Behroozi (2021, in preparation). 


\def\figsize{0.65}
\begin{figure*}
\centering
\includegraphics[trim =  11mm 30mm 6mm 5mm, clip, scale=\figsize]{\figdir/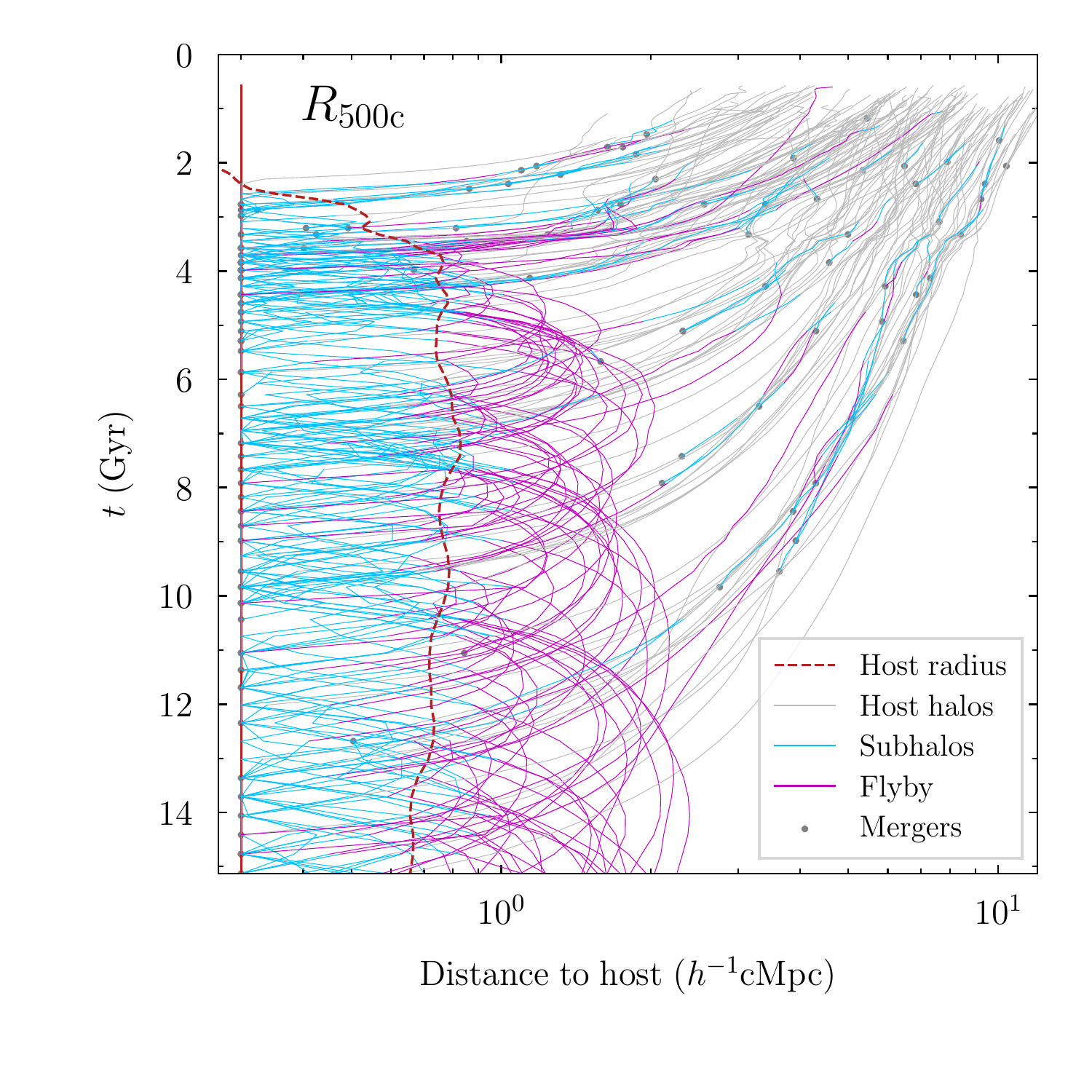}
\includegraphics[trim =  30mm 30mm 6mm 5mm, clip, scale=\figsize]{\figdir/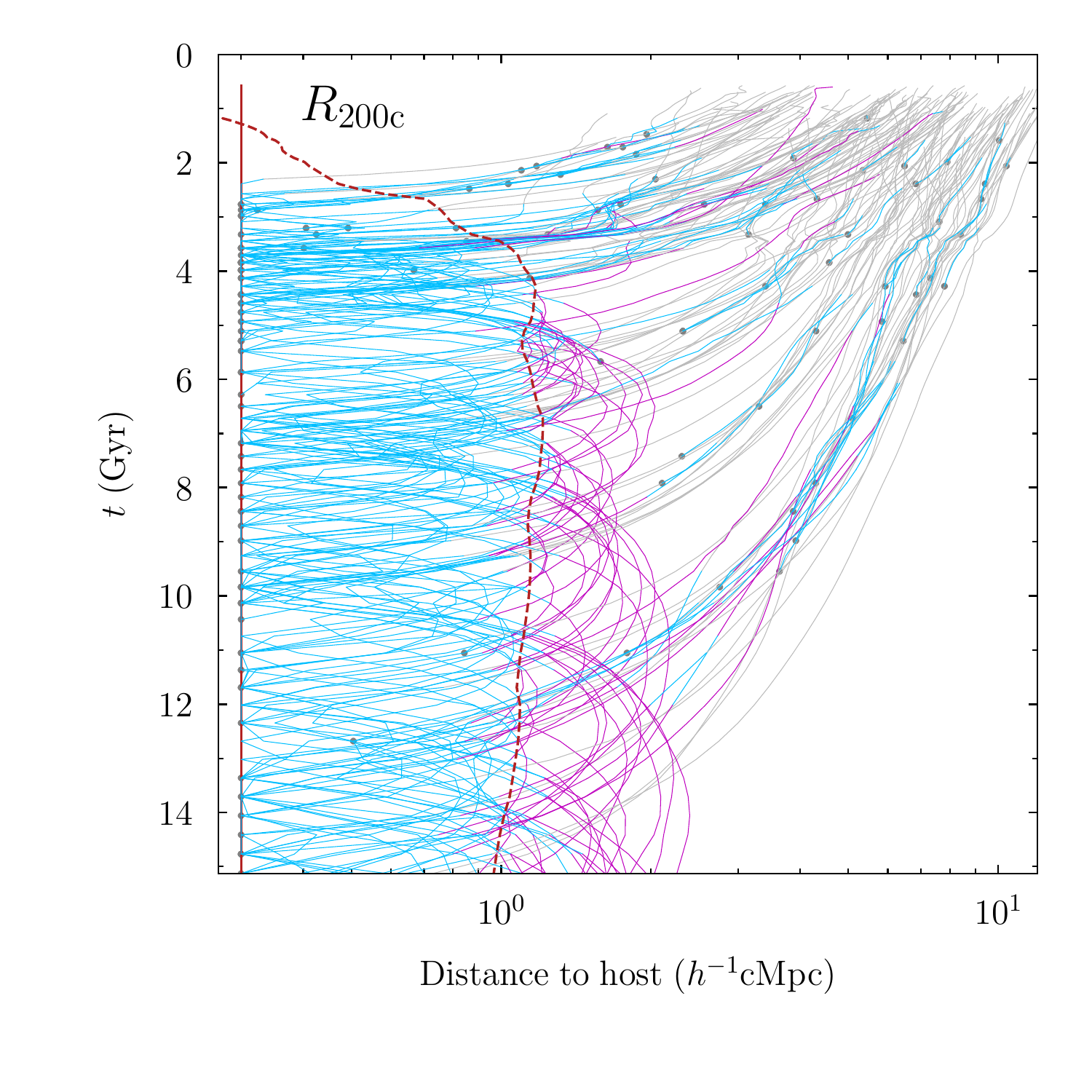}
\includegraphics[trim =  11mm 12mm 6mm 5mm, clip, scale=\figsize]{\figdir/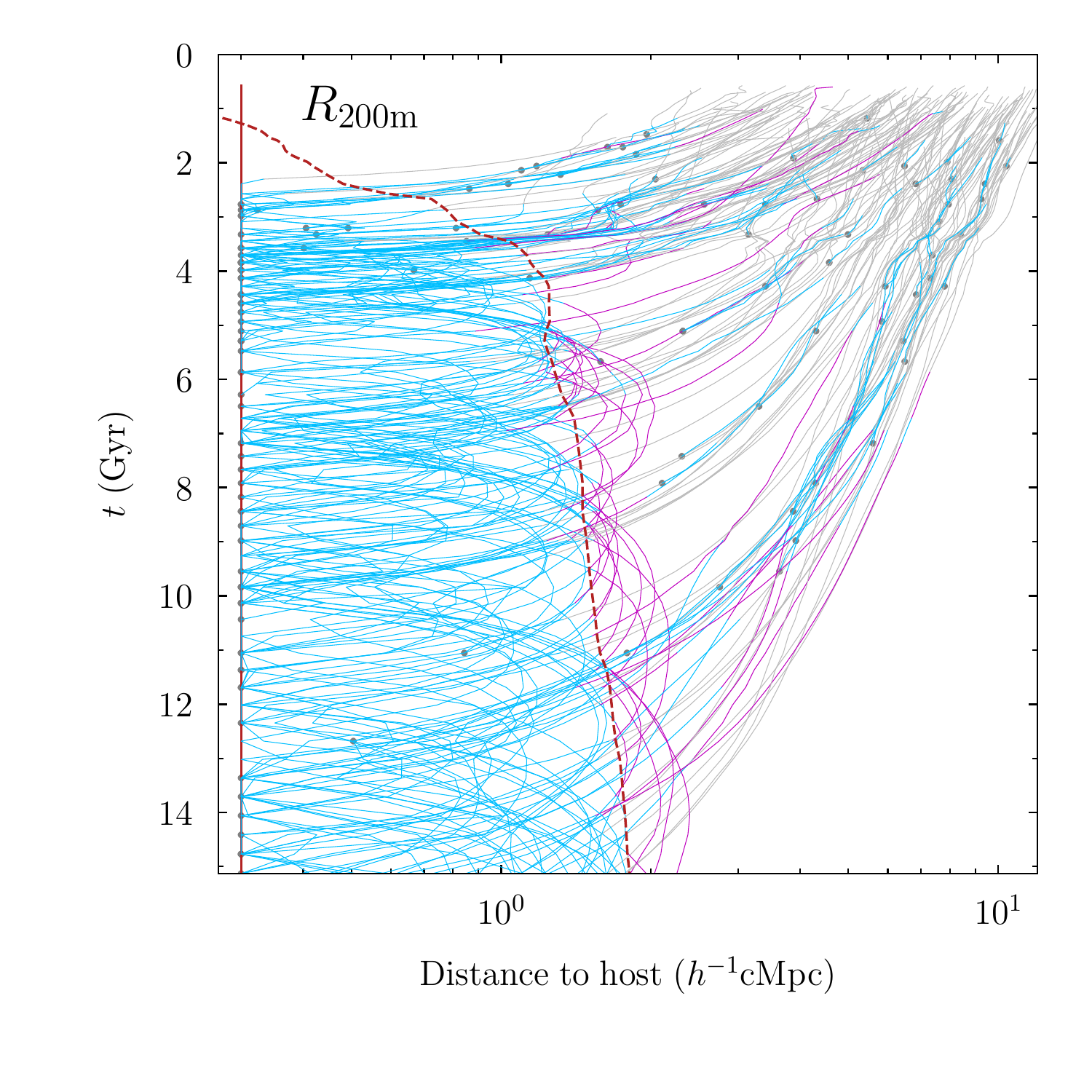}
\includegraphics[trim =  30mm 12mm 6mm 5mm, clip, scale=\figsize]{\figdir/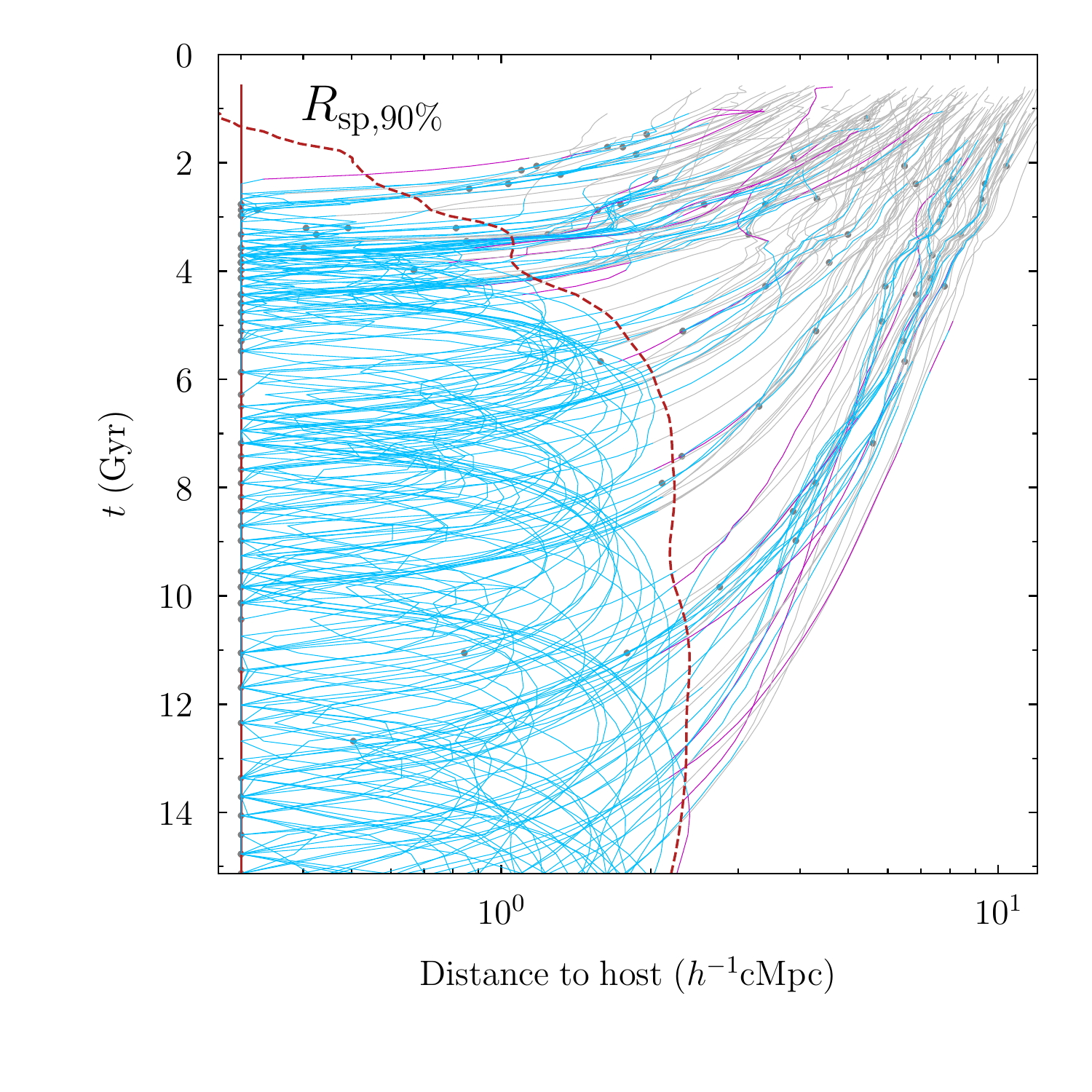}
\caption{Visualization of a merger tree based on different mass definitions. Each panel shows the history of the largest halo in TestSim100. The trajectories reflect the distance from this ``root'' halo in comoving units. The distance is arbitrarily cut off at a radius well inside the root halo, represented by the vertical line on the left of the plots. Each track represents a halo that becomes a subhalo of the root halo, merges into it, or merges into one of its subhalos. The color of the lines indicates the epochs when those halos are hosts (gray), subhalos (of any halo; light blue), or flyby halos (purple). The radius definition chosen to define the host--subhalo relations changes both the halos included in the tree and their status. Small radii such as the commonly used $\rtoc$ produce a large number of flyby halos whereas the splashback radius includes virtually all subhalos by construction.}
\label{fig:trees}
\end{figure*}

\begin{deluxetable*}{lcccccccccccc}
\tablecaption{$N$-body Simulations
\label{table:sims}}
\tablewidth{\textwidth}
\tablehead{
\colhead{Name} &
\colhead{$L\, (\mpch)$} &
\colhead{$N^3$} &
\colhead{$m_{\rm p}\, (\msunh)$} &
\colhead{$\epsilon\, (\kpch)$} &
\colhead{$\epsilon / (L / N)$} &
\colhead{$z_{\rm initial}$} &
\colhead{$z_{\rm final}$} &
\colhead{$N_{\rm snaps}$} &
\colhead{$z_{\rm f-snap}$} &
\colhead{$z_{\rm f-cat}$} &
\colhead{Cosmology} &
\colhead{Reference}
}
\startdata
L2000-WMAP7 & $2000$ & $1024^3$ & $5.6 \times 10^{11}$  & $65$  & $1/30$ & $49$ & $0$ & $100$ & $20$ & $4.2$ & WMAP7 & \citetalias{diemer_15} \\
L1000-WMAP7 & $1000$ & $1024^3$ & $7.0 \times 10^{10}$ & $33$ & $1/30$ & $49$ & $0$ &  $100$ & $20$ & $6.2$ & WMAP7 & \citetalias{diemer_13_scalingrel} \\
L0500-WMAP7 & $500$  & $1024^3$ & $8.7 \times 10^{9}$  & $14$ & $1/35$  & $49$ & $0$ &  $100$ & $20$ & $8.8$ & WMAP7 & \citetalias{diemer_14} \\
L0250-WMAP7 & $250$  & $1024^3$ & $1.1 \times 10^{9}$  & $5.8$  & $1/42$  & $49$ & $0$ &  $100$ & $20$ & $11.5$ & WMAP7 & \citetalias{diemer_14} \\
L0125-WMAP7 & $125$  & $1024^3$ & $1.4 \times 10^{8}$  & $2.4$  & $1/51$  & $49$ & $0$ &  $100$ & $20$ & $14.5$ & WMAP7 & \citetalias{diemer_14} \\
L0063-WMAP7 & $62.5$ & $1024^3$ & $1.7 \times 10^{7}$  & $1.0$  & $1/60$ & $49$ & $0$ &  $100$ & $20$ & $17.6$ & WMAP7 & \citetalias{diemer_14} \\
L0031-WMAP7 & $31.25$ & $1024^3$ & $2.1 \times 10^{6}$  & $0.25$  & $1/122$ & $49$ & $2$ &  $64$ & $20$ & $20$ & WMAP7 & \citetalias{diemer_15} \\
TestSim100 & $62.5$  & $256^3$  & $1.1 \times 10^{9}$  & $5.8$  & $1/42$  & $49$ & $-0.1$ & $96$  & $9$  & $9$ & WMAP7 & \citetalias{diemer_17_sparta} \\
L0500-Planck & $500$  & $1024^3$ & $1.0 \times 10^{10}$  & $14$ & $1/35$  & $49$ & $0$ &  $100$ & $20$ & $9.1$ & \planck & \citetalias{diemer_15} \\
L0250-Planck & $250$  & $1024^3$ & $1.3 \times 10^{9}$  & $5.8$  & $1/42$  & $49$ & $0$ &  $100$ & $20$ & $12.3$ & \planck & \citetalias{diemer_15} \\
L0125-Planck & $125$  & $1024^3$ & $1.6 \times 10^{8}$  & $2.4$  & $1/51$  & $49$ & $0$ &  $100$ & $20$ & $15.5$ & \planck & \citetalias{diemer_15} \\
L0100-PL-1.0 & $100$  & $1024^3$ & $2.6 \times 10^{8}$  & $0.5$  & $1/195$  & $119$ & $2$ & $64$ & $20$ & $20$ & PL, $n=-1.0$ & \citetalias{diemer_15} \\
L0100-PL-1.5 & $100$  & $1024^3$ & $2.6 \times 10^{8}$  & $0.5$  & $1/195$  & $99$ & $1$ & $78$ & $20$ & $20$ & PL, $n=-1.5$ & \citetalias{diemer_15} \\
L0100-PL-2.0 & $100$  & $1024^3$ & $2.6 \times 10^{8}$  & $1.0$  & $1/98$  & $49$ & $0.5$ & $100$ & $20$ & $15.5$ & PL, $n=-2.0$ & \citetalias{diemer_15} \\
L0100-PL-2.5 & $100$  & $1024^3$ & $2.6 \times 10^{8}$  & $1.0$  & $1/98$  & $49$ & $0$ & $100$ & $20$ & $5.4$ & PL, $n=-2.5$ & \citetalias{diemer_15} 
\enddata
\tablecomments{The $N$-body simulations used in this paper. $L$ denotes the box size in comoving units, $N^3$ the number of particles, $m_{\rm p}$ the particle mass, and $\epsilon$ the force softening length in physical units. The simulations cover redshifts from $z_{\rm initial}$ to $z_{\rm final}$, but snapshots were output only between $z_{\rm f-snap}$ and $z_{\rm final}$; the catalogs contain the first halos at $z_{\rm f-cat}$. The references correspond to \citet[][\citetalias{diemer_13_scalingrel}]{diemer_13_scalingrel}, \citet[][\citetalias{diemer_14}]{diemer_14}, and \citet[][\citetalias{diemer_15}]{diemer_15}. Our system for choosing force resolutions is discussed in \citetalias{diemer_14}.}
\end{deluxetable*}

\section{Simulation Data}
\label{sec:method}

In this section, we briefly review our simulations, radius definitions, and our algorithm for assigning subhalos to hosts. We refer the reader to \citetalias{diemer_20_catalogs} for details.

\subsection{Simulations and Halo Catalogs}
\label{sec:method:data}

Our catalogs are based on the \erebos suite of dissipationless $N$-body simulations, which we summarize in Table~\ref{table:sims}. This suite includes seven simulations of a \wmap \LCDM cosmology based on that of the Bolshoi simulation \citep[][$\Omega_{\rm m} = 0.27$, $\Omega_{\rm b} = 0.0469$, $h = 0.7$, $\sigma_8 = 0.82$, and $n_{\rm s} = 0.95$]{klypin_11, komatsu_11}. The simulations span a range from $31 \mpch$ to $2000 \mpch$ in box size and allow us to investigate a large range of halo masses.  We also use three simulations of a \planck-like cosmology \citep[][$\Omega_{\rm m} = 0.32$, $\Omega_{\rm b} = 0.0491$, $h = 0.67$, $\sigma_8 = 0.834$, and $n_{\rm s} = 0.9624$]{planck_14}. The \planck results are virtually identical to those from the \wmap cosmology and are only used to constrain our fitting function. We omit them from the figures to avoid crowding and emphasize that ``\LCDM'' labels refer to the \wmap simulations. We also consider self-similar Einstein--de Sitter universes with power-law initial power spectra of slopes $-1$, $-1.5$, $-2$, and $-2.5$ \citepalias{diemer_20_catalogs}. These simulations allow us to test the dependence of our results on the initial power spectrum \citep[e.g.,][]{efstathiou_88, elahi_09_plsubs, brown_20, joyce_20}. The initial power spectra for the \LCDM simulations were computed by \textsc{Camb} \citep{lewis_00}. The initial particle grids for all simulations were generated by \textsc{2LPTic} \citep{crocce_06}, and the simulations were run with \textsc{Gadget2} \citep{springel_05_gadget2}.

All figures in this work are based on the merger trees presented in \citetalias{diemer_20_catalogs}. We start from halo catalogs and merger trees created with \rockstar and \consistenttrees \citep{behroozi_13_rockstar, behroozi_13_trees}. We then run the \sparta code on each simulation to measure splashback radii and other halo properties \citepalias{diemer_17_sparta}. The \moria extension recombines the \sparta results with the \rockstar catalogs to create enhanced catalogs and merger trees in a new format \citepalias{diemer_20_catalogs}.

\subsection{Definitions of Halo Radius and Mass}
\label{sec:method:defs}

We use three types of radius definitions: bound-only SO radii, all-particle SO radii, and splashback radii. \rockstar calculates bound-only radii by removing gravitationally unbound particles from friends-of-friends groups and subgroups in six-dimensional phase space. Given a set of bound particles, SO mass definitions are computed by finding the outermost radius where the density falls below the given SO threshold. We consider four definitions, $\rfoc$, $\rtoc$, $\rvir$, and $\rtom$, indicating density thresholds of $500$ or $200$ times the critical or mean density of the universe. We compute the varying virial overdensity using the approximation of \citet{bryan_98}. Second, we consider all-particle SO radii computed by \sparta, which are measured the same way as the bound-only radii but without any unbinding. For the vast majority of host halos, the difference is small, but for subhalos, all-particle masses are ill-defined because they often contain large amounts of host material \citepalias{diemer_20_catalogs}. In this work, we are mostly concerned with host--subhalo relations, meaning that the exact mass of a subhalo does not matter as much as the radius of its host. Nevertheless, we will mostly rely on bound-only radii and show results for the all-particle virial radius for comparison. For brevity, we denote the bound-only and all-particle versions of a definition $X$ as $R_{\rm X,bnd}$ and $R_{\rm X,all}$.

Finally, we use splashback radii computed by \sparta. The code tracks each particle in each halo as it enters for the first time and determines its first apocenter (or splashback) event. From the locations and times of these events, we compute the splashback radius of the halo by smoothing the distribution of particle splashbacks in time and taking its mean ($R_{\rm sp,mn}$) or higher percentiles (e.g., $R_{\rm sp,90\%}$ for the 90th percentile). At the final snapshots of a simulation, the time average would be biased because we are missing future particle splashbacks. \sparta corrects for this bias, but this procedure increases the scatter in the splashback results for the final few snapshots \citepalias{diemer_17_sparta}. Thus, we will study results at $z = 0.13$ instead of $z = 0$, which does not alter our conclusions in any way.

Any mass $M_{\rm X}$ is understood to include the mass inside the respective radius $R_{\rm X}$, and $N_{\rm X}$ denotes the number of particles in $M_{\rm X}$. The peak mass $M_{\rm X,peak}$ is the highest mass attained along a halo's most-massive progenitor branch. The \moria catalogs and merger trees contain all halos with $N_{\rm 200m,peak} \geq 200$, but we will generally apply stricter limits to avoid selection effects (Appendix~\ref{sec:app}). Finally, we consider two alternative ways to quantify the relative masses of halos. First, we use the maximum circular velocity, $\vmax$, as computed by \rockstar. Second, when comparing halos across redshifts and cosmologies, we express their masses as peak height, $\nu_{\rm X}$. Peak height captures the statistical significance of halos, namely, whether they are rare or common with respect to the overall density field. It is formally defined as $\nu_{\rm X} = \deltac / \sigma(M_{\rm X})$, where $\deltac(z) = 1.686 \times \Omega_{\rm m}(z)^{0.0055}$ is the threshold overdensity in the top-hat collapse model \citep{gunn_72, mo_10_book} and $\sigma(M_{\rm X})$ is the variance of the linear power spectrum. We include a correction for the finite volume of our simulations (see \citetalias{diemer_20_catalogs} or \citealt{diemer_20_mfunc} for details). The variance is measured on a scale of the Lagrangian radius of a halo, $R_{\rm L}$, which corresponds to the comoving radius that encloses the mass $M_{\rm X}$ at the mean density of the universe,
\begin{equation}
\label{eq:rl}
M_{\rm L} = M_{\rm X} = (4 \pi/3) \rho_{\rm m}(z=0) R_{\rm L}^3 \,.
\end{equation}
We can compute this radius for any mass definition $M_{\rm X}$, but we mostly use $\nutom$. We compute peak heights with the \colossus code \citep{diemer_18_colossus}, using the transfer function of \citet{eisenstein_98} to approximate the power spectrum.

\subsection{Host--Subhalo Relations}
\label{sec:method:hostsub}

One of the main innovations of our \moria catalogs is that they contain separate host--subhalo relations for each definition. To compute these relations, \moria orders the list of all halos at a snapshot by $\vmax$ to avoid making reference to any particular mass definition. Starting with the highest-$\vmax$ halo, the code searches for all halo centers within its radius in the given definition and assigns them the host's ID as a parent. We then continue with the second-highest $\vmax$ and so on. If a subhalo already has a host, we do not replace that host's ID. This procedure exactly reproduces the parent assignments of \consistenttrees if the same radius definition is used. Figure~\ref{fig:viz} shows a visualization of the host--subhalo assignments for different mass definitions. The details of the percolation algorithm have some impact on the results \citep{garcia_19}, but these differences are not the subject of this paper and are small compared to the changes caused by varying the size of the halo radius.


\section{Results}
\label{sec:results}

In this section, we quantify the impact of the radius definition on the subhalo and flyby fractions in \LCDM (Sections~\ref{sec:results:subfrac} and \ref{sec:results:flyby}). We compare the fractions across cosmologies and show that they follow a universal form (Section~\ref{sec:results:universality}), for which we provide a simple fitting function (Section~\ref{sec:results:fit}).

\subsection{Subhalo Fraction}
\label{sec:results:subfrac}

\def\figsize{0.61}
\begin{figure*}
\centering
\includegraphics[trim =  3mm 23mm 1mm 2mm, clip, scale=\figsize]{\figdir/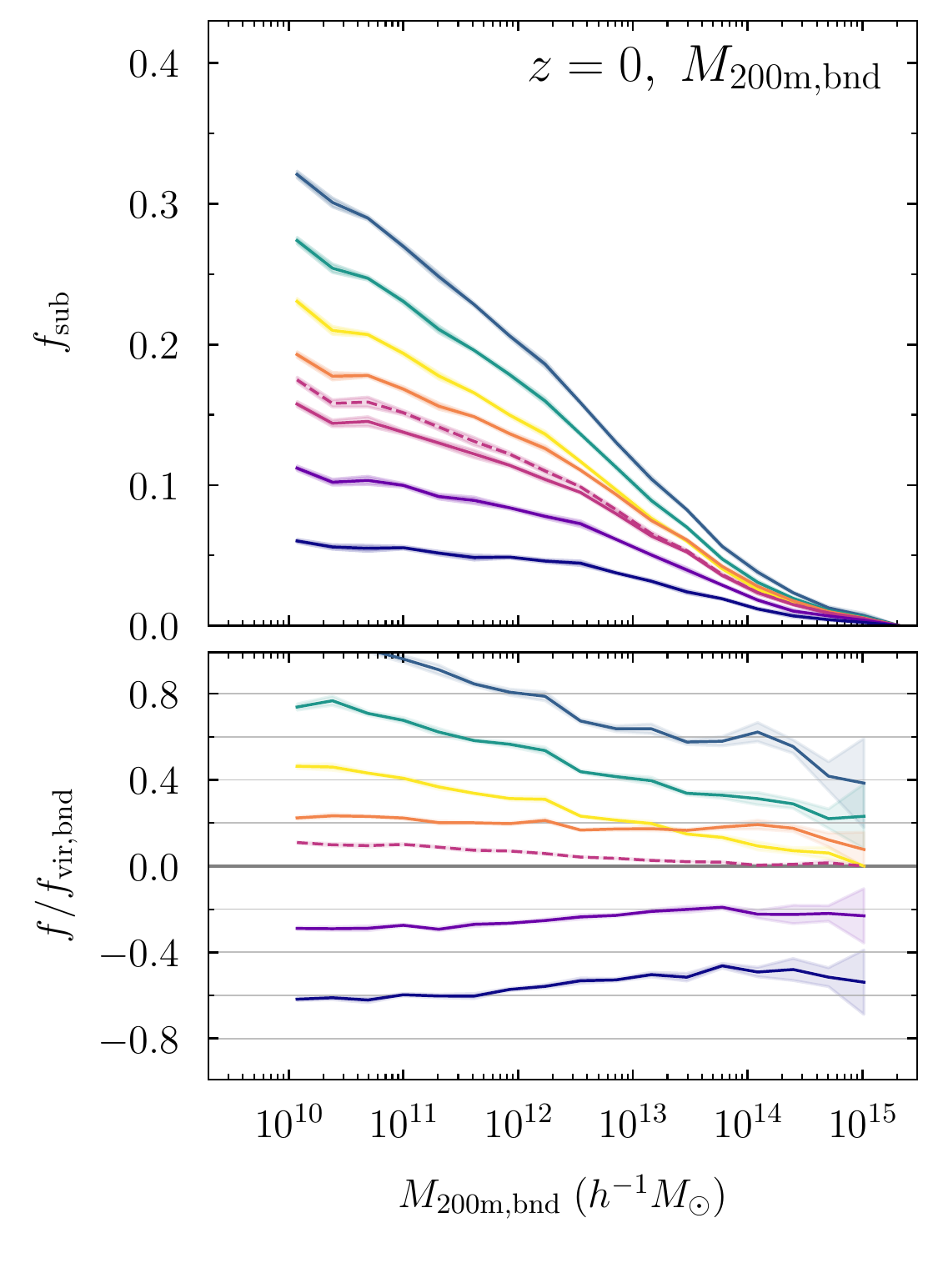}
\includegraphics[trim =  25mm 23mm 1mm 2mm, clip, scale=\figsize]{\figdir/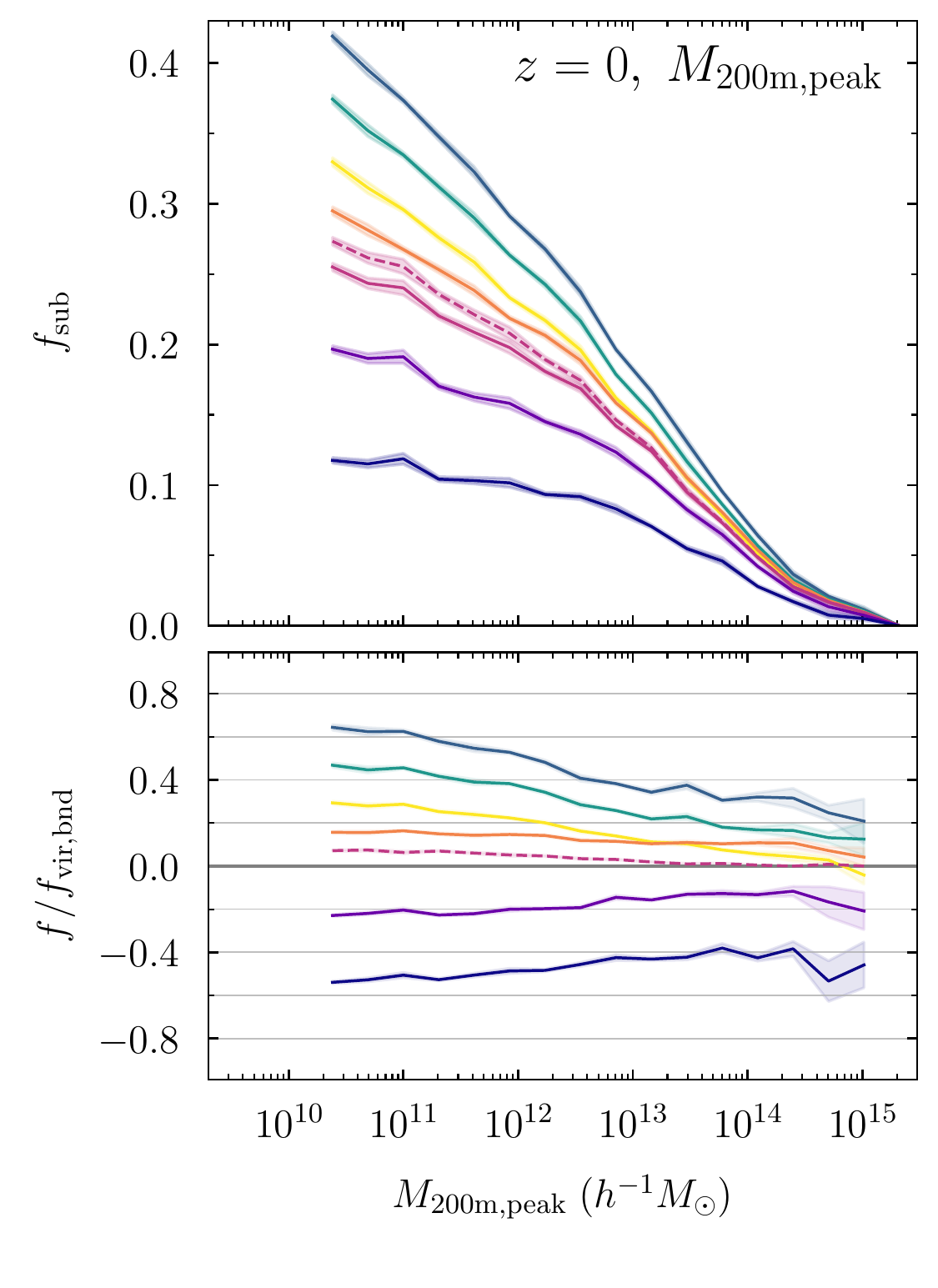}
\includegraphics[trim =  25mm 23mm 1mm 2mm, clip, scale=\figsize]{\figdir/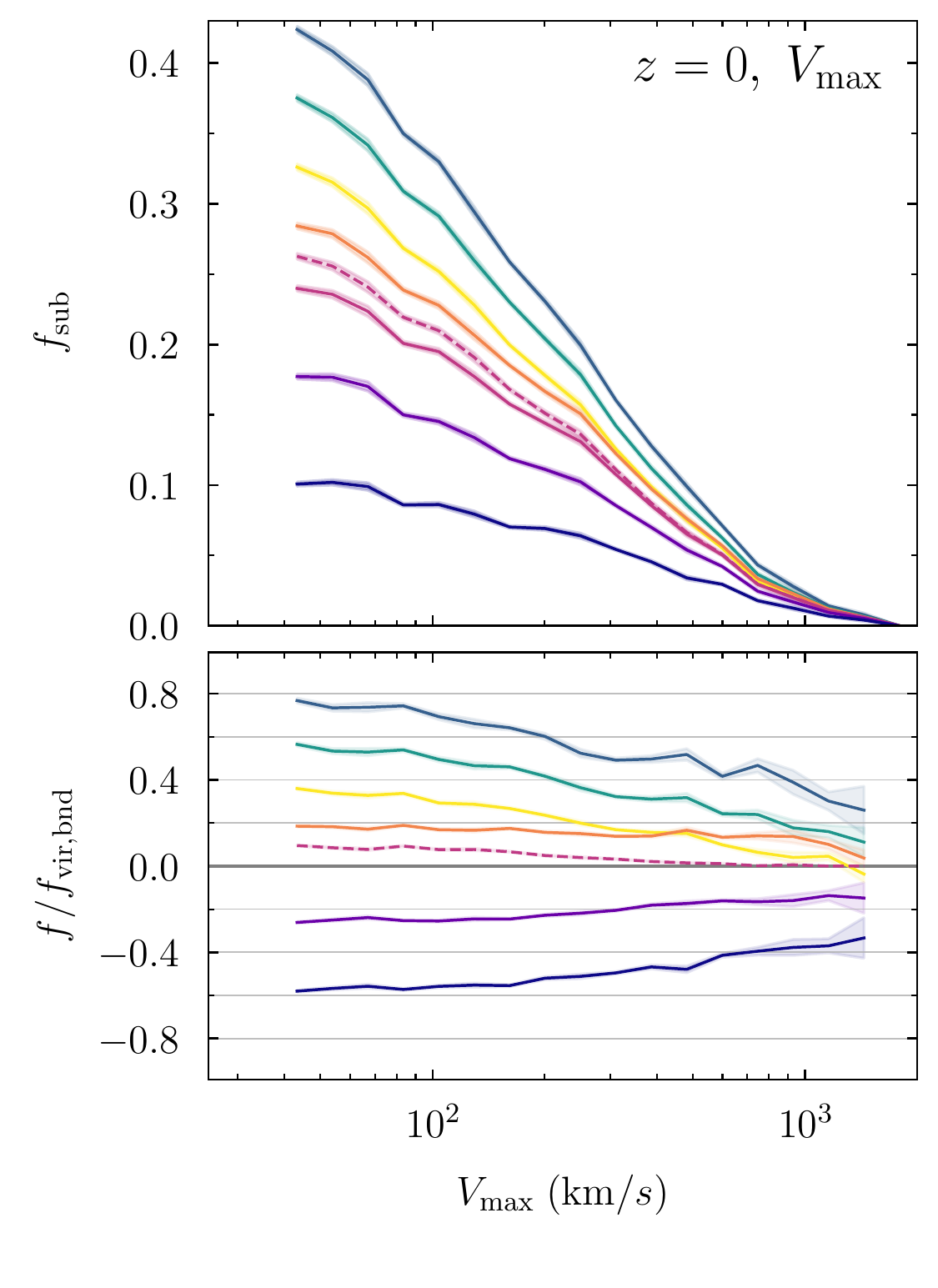}
\includegraphics[trim =  3mm 7mm 1mm 2mm, clip, scale=\figsize]{\figdir/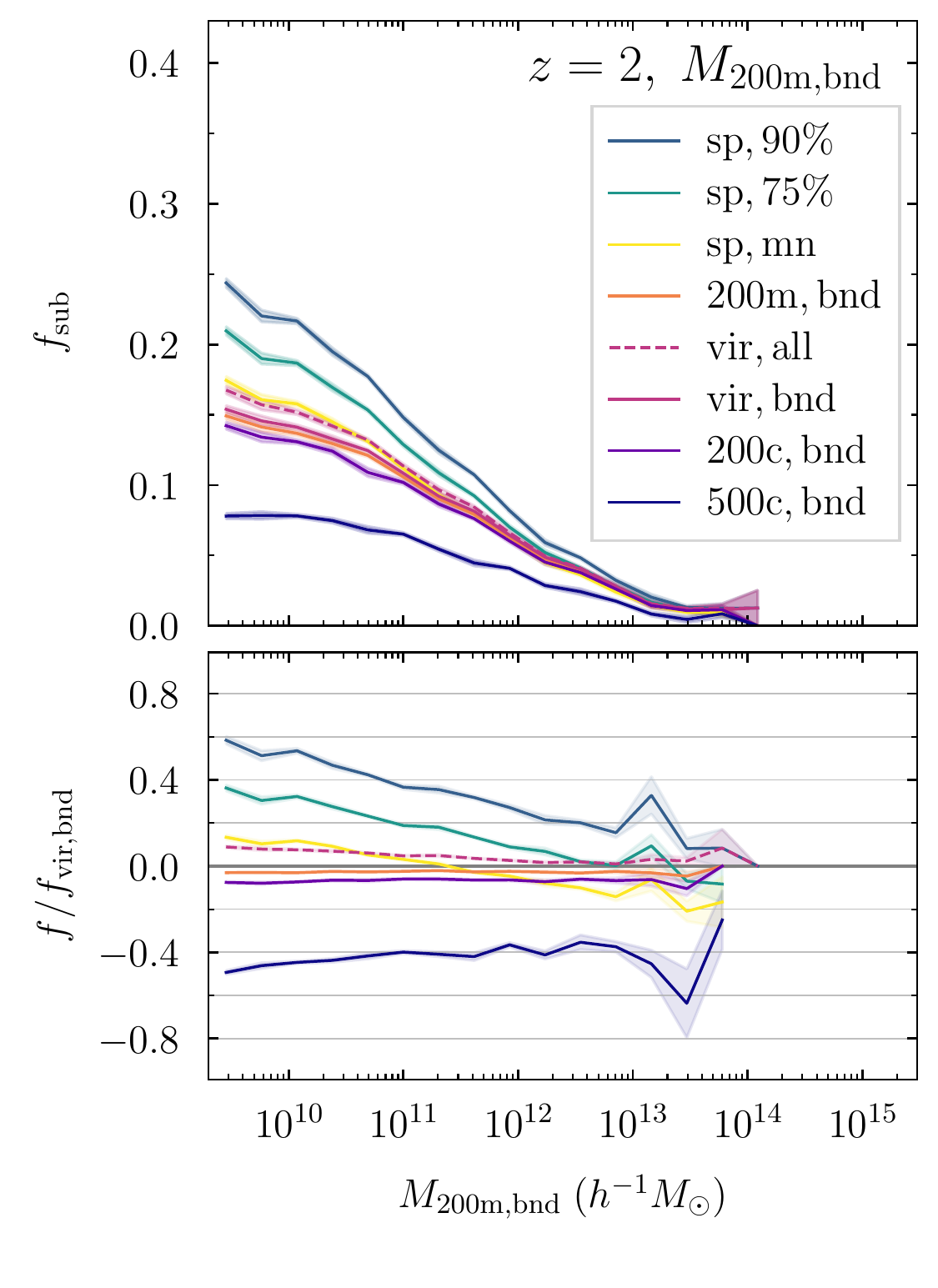}
\includegraphics[trim =  25mm 7mm 1mm 2mm, clip, scale=\figsize]{\figdir/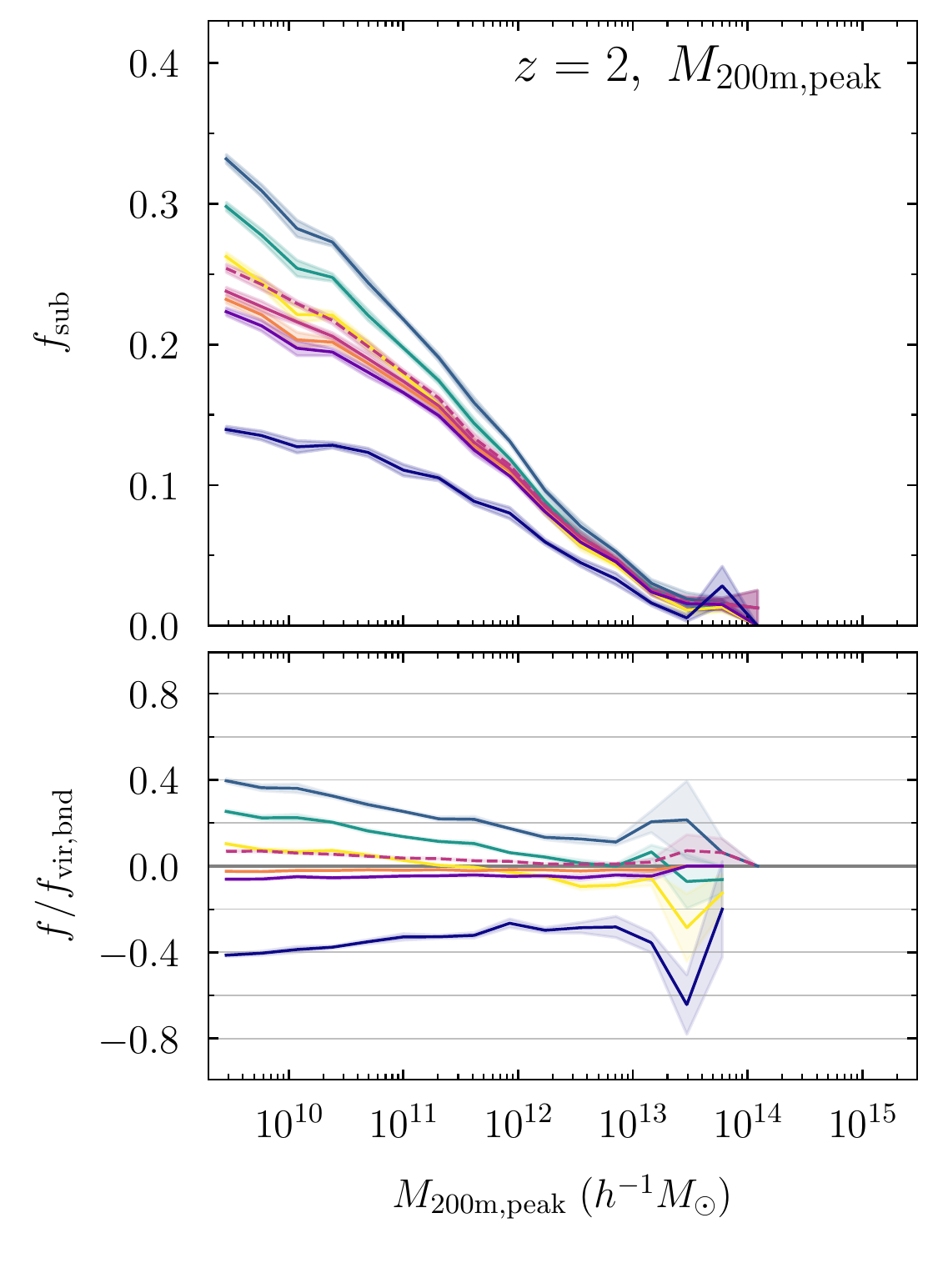}
\includegraphics[trim =  25mm 7mm 1mm 2mm, clip, scale=\figsize]{\figdir/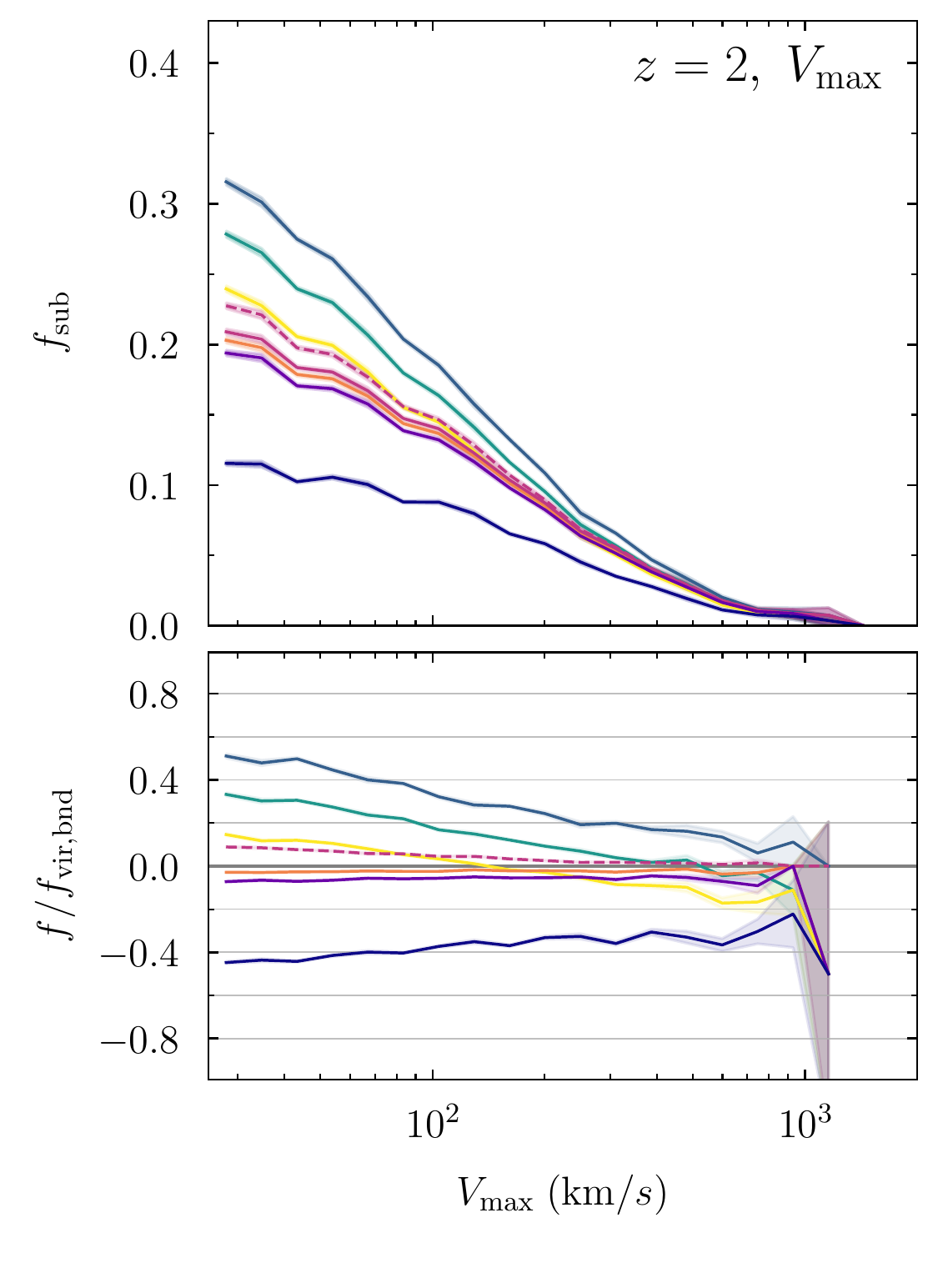}
\caption{Subhalo fraction in \LCDM according to different mass definitions as a function of the current bound-only halo mass (left), the peak mass (center), and $\vmax$ (right). These definitions cause significant shifts in the inferred subhalo fractions because subhalos tend to lose mass after infall whereas host halos keep growing. The top row shows the relations at $z \approx 0$ (left), the bottom row at $z = 2$. The smaller bottom panels show the fractional difference to the commonly used $R_{\rm vir,bnd}$ definition. The shaded areas highlight the statistical uncertainty. The dashed line refers to $R_{\rm vir,all}$, which is similar to $R_{\rm vir,bnd}$; we omit the all-particle counterparts of the other SO definitions. At fixed $\vmax$ and $z = 0$, using $\rfoc$ leads to about 60\% fewer subhalos at the low-mass end, $\rtom$ to about 20\% more. The three representative splashback definitions lead to between 30\% and 80\% higher subhalo fractions at the low-mass end and up to 50\% at cluster masses. All differences are reduced at higher redshift, mostly reflecting the overall shift to lower masses. For the most-massive halos at $z \geq 2$, the average mass accretion rates are so high that the smaller splashback radii such as $R_{\rm sp,mn}$ lie inside $\rvir$ on average. The virial, 200c, and 200m definitions become indistinguishable at high redshift. These results highlight the degree to which the definition of the halo boundary affects our understanding of substructure.}
\label{fig:frac}
\end{figure*}

\def\figsize{0.61}
\begin{figure*}
\centering
\includegraphics[trim =  0mm 23mm 1mm 2mm, clip, scale=\figsize]{\figdir/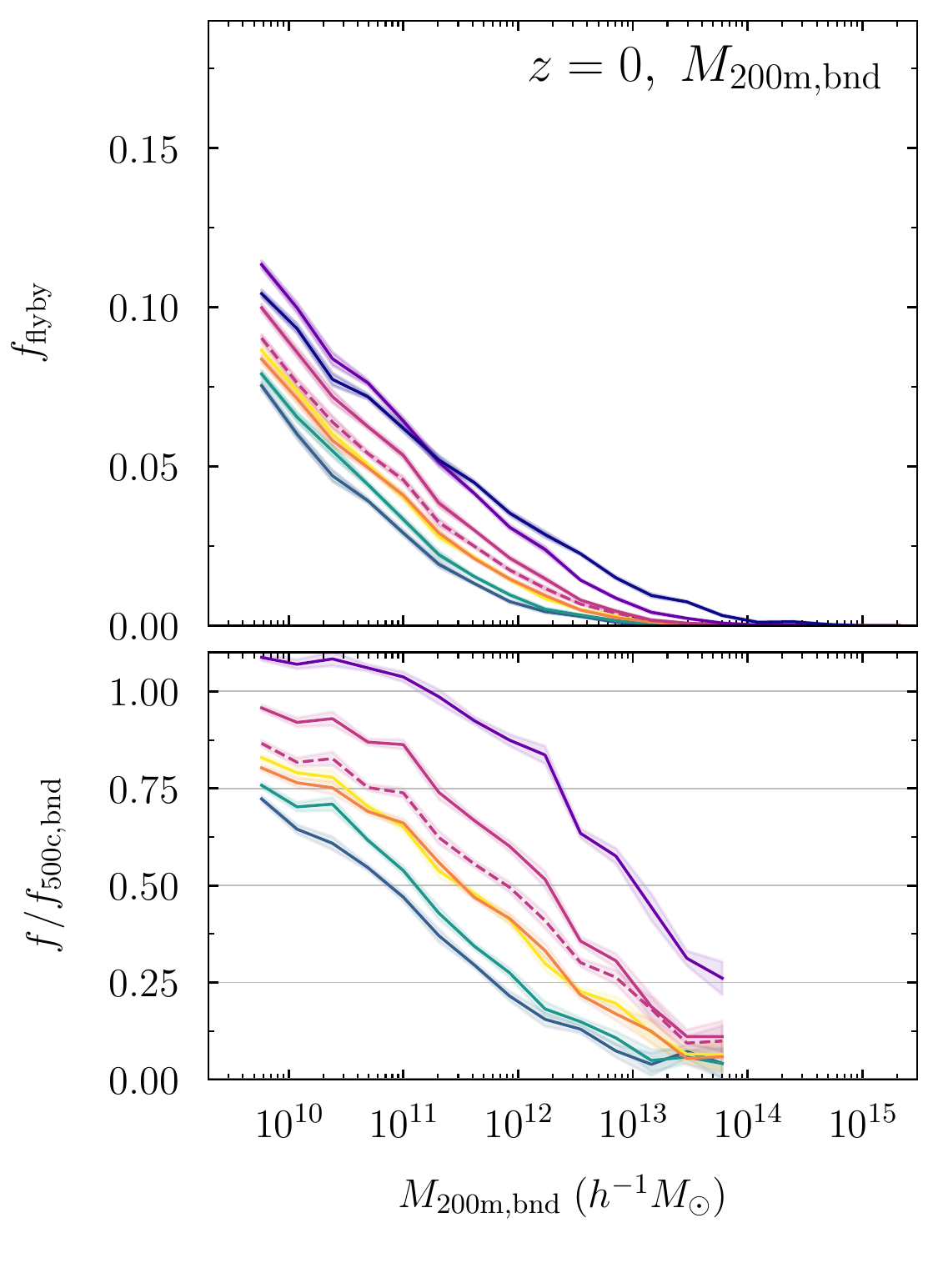}
\includegraphics[trim =  25mm 23mm 1mm 2mm, clip, scale=\figsize]{\figdir/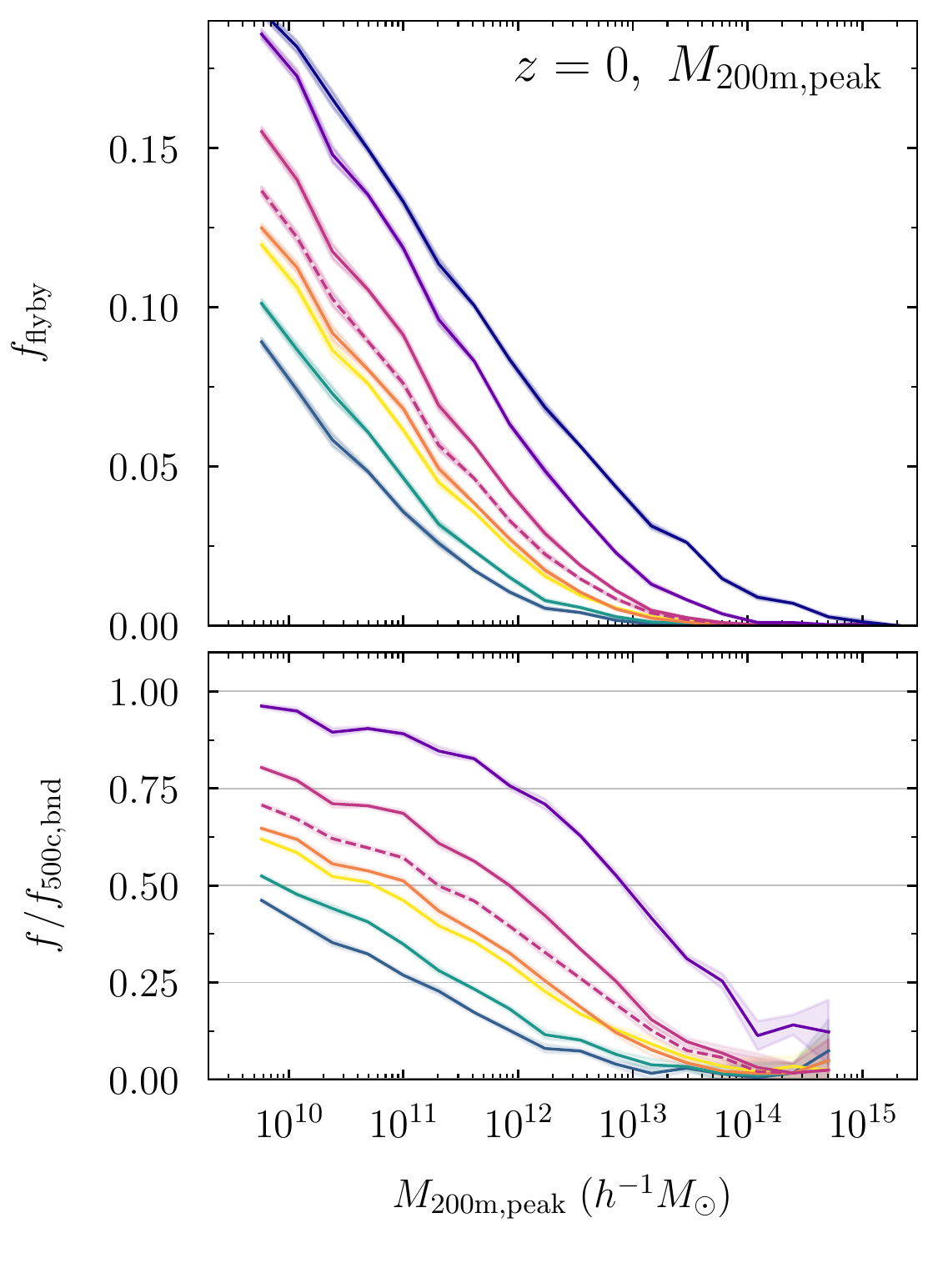}
\includegraphics[trim =  25mm 23mm 1mm 2mm, clip, scale=\figsize]{\figdir/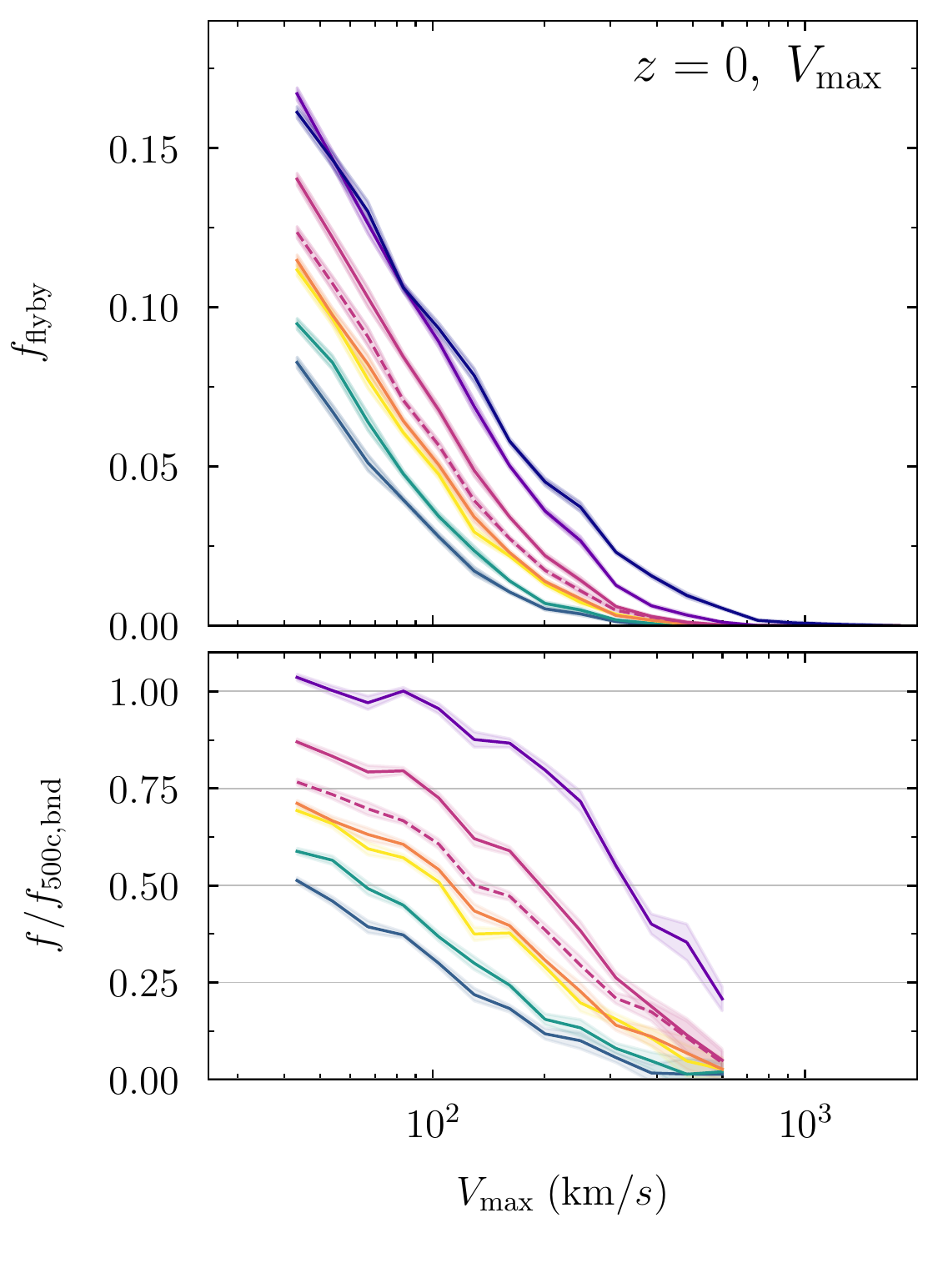}
\includegraphics[trim =  0mm 7mm 1mm 2mm, clip, scale=\figsize]{\figdir/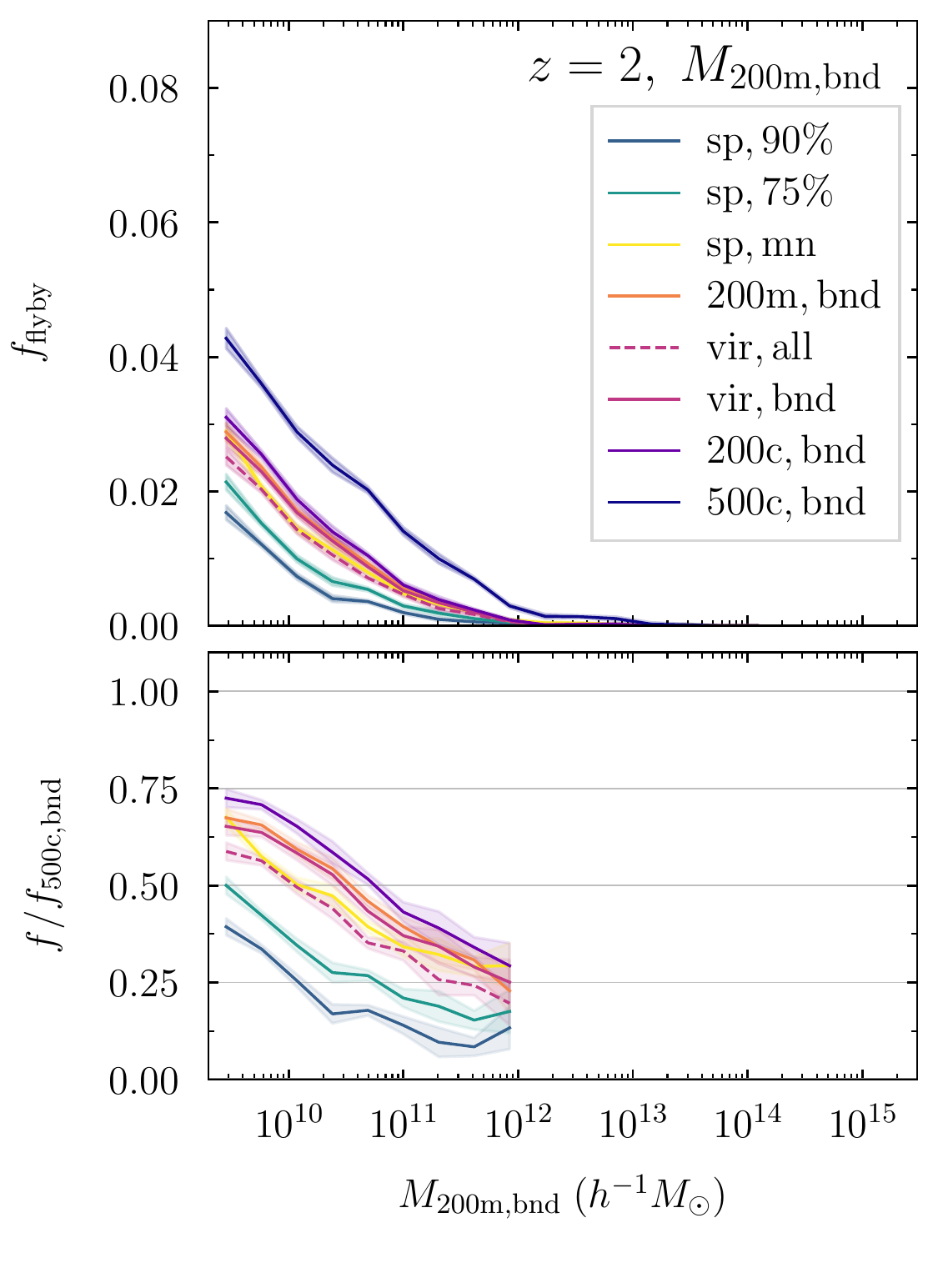}
\includegraphics[trim =  25mm 7mm 1mm 2mm, clip, scale=\figsize]{\figdir/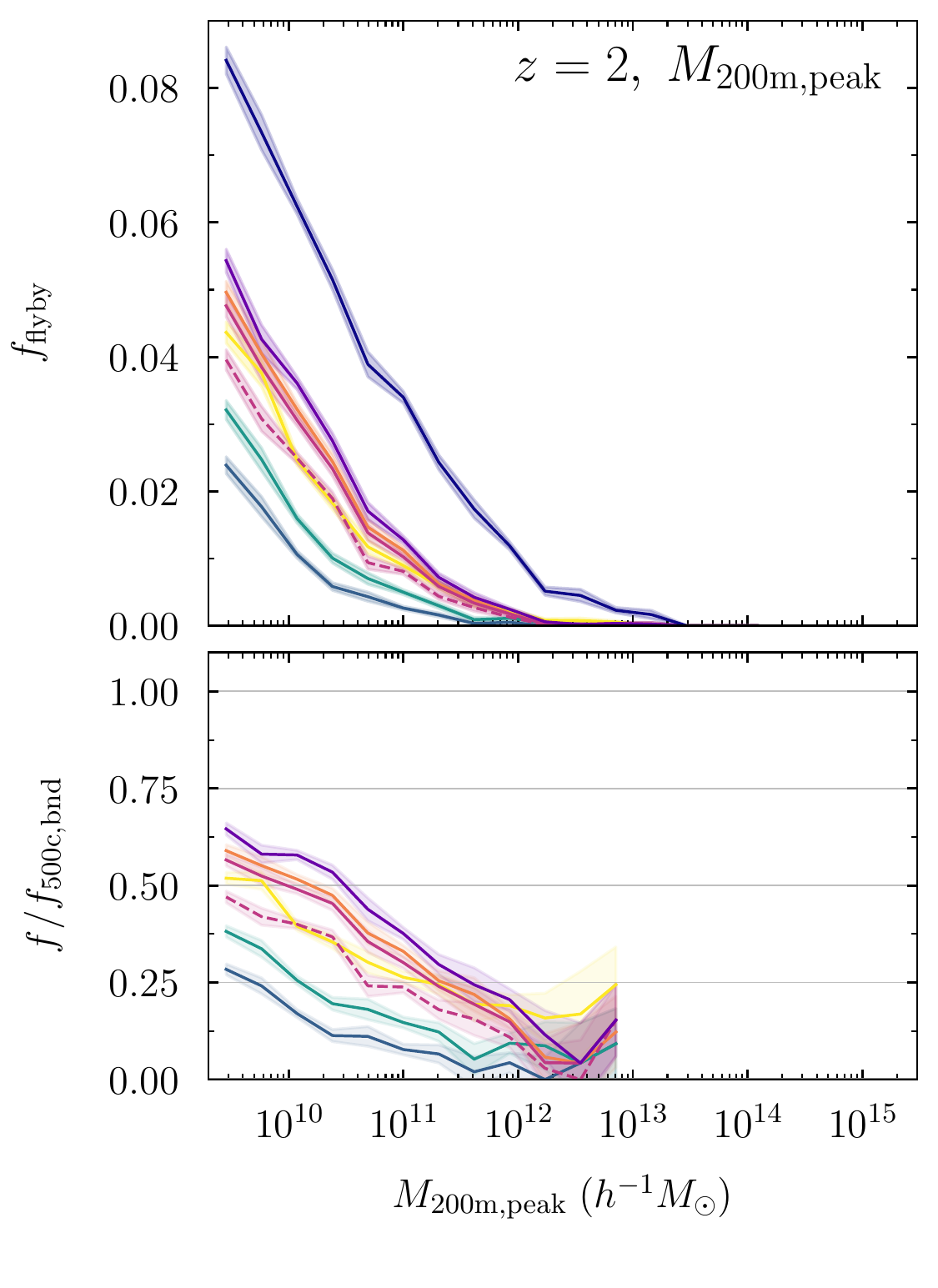}
\includegraphics[trim =  25mm 7mm 1mm 2mm, clip, scale=\figsize]{\figdir/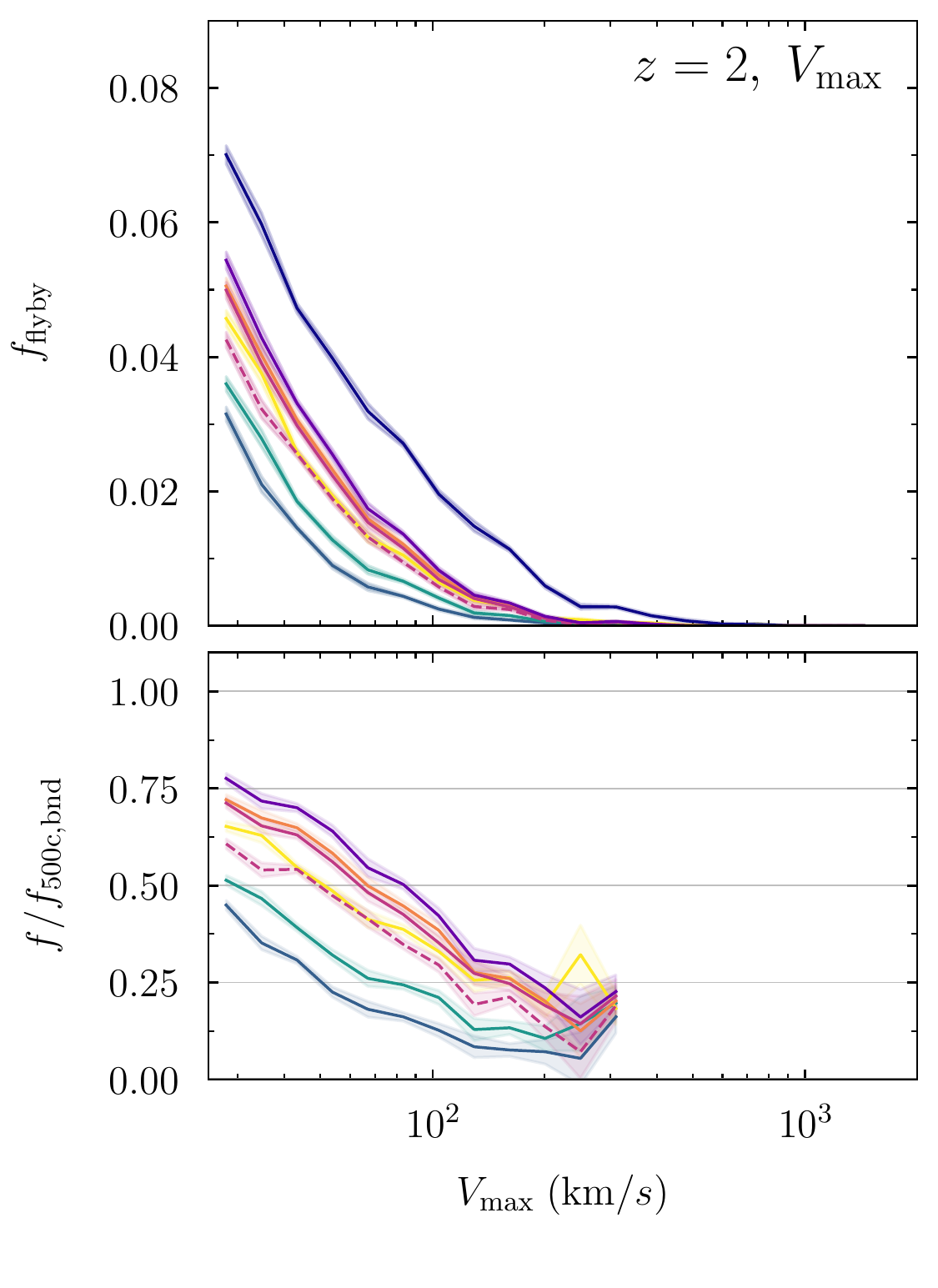}
\caption{Same as Figure~\ref{fig:frac} but for the flyby fraction, which also depends strongly on the radius definition. Conversely, though, small-radius definitions such as $\rfoc$ (dark blue) produce many more flyby halos than large-radius definitions. The bottom panels compare all definitions to $\rfoc$ (excluding masses where the fraction approaches zero). See Section~\ref{sec:results:flyby} for a detailed discussion.}
\label{fig:bs}
\end{figure*}

We define the subhalo fraction, $\fsub$, as the fraction of all halos with a given mass (or similar characteristic) that are subhalos. At first sight, $\fsub$ should be easy to measure by counting the number of halos that do and do not have a parent at fixed mass. The difficulty is to define a halo property that can be meaningfully measured for both host and subhalos. As briefly discussed in Section~\ref{sec:method:defs} (and at length in \citetalias{diemer_20_catalogs}), all-particle SO masses are ill-defined for subhalos, but we can use bound-only masses, peak masses, or $\vmax$. Because each choice leads to different subhalo fractions, Figure~\ref{fig:frac} shows $\fsub$ based on all three quantities. Here, we combine halos from all \wmap simulations as long as they satisfy certain resolution criteria; we do not show the individual simulations to avoid crowding. In the smaller bottom panels of Figure~\ref{fig:frac}, we compare the subhalo fractions in each definition to the commonly used $R_{\rm vir,bnd}$. The uncertainties are computed using jackknife resampling. We omit ratios of zero and very noisy measurements for clarity. The details of our procedure, resolution limits, and numerical convergence are described in the Appendix.

First, we consider the subhalo fraction as a function of the current bound-only mass $M_{\rm 200m,bnd}$. Overall, $\fsub$ monotonically decreases from between 6\% and 32\% at the low-mass end to zero at the highest masses. The differences between the mass definitions are striking: compared to $\rvir$, high-threshold SO definitions such as $\rtoc$ reduce the subhalo fraction by up to 50\%, $\rtom$ increases it by about 20\%, and the splashback definitions increase it by 20--40\% for the mean and 50--120\% for the 90th percentile. At $z = 2$, $\rtoc$, $\rvir$, and $\rtom$ have become almost indistinguishable because they approach the same overdensity threshold when $\Omega_{\rm m}(z) \approx 1$. $\rfoc$ still reduces the subhalo fraction by a large factor, although slightly less than at $z = 0$. The splashback definitions lead to somewhat smaller increases in the subhalo fraction because the higher accretion rates at high redshift mean that they shrink compared to $\rtom$ \citepalias{diemer_17_rsp}. Using $R_{\rm vir,all}$ instead of $R_{\rm vir,bnd}$ makes a relatively small difference, up to 10\% at the low-mass end. The differences for other SO thresholds are similar so that our conclusions for bound-only definitions basically apply to their all-particle counterparts as well. This result is congruent with \citet{diemer_20_mfunc}, who show that the mass function of all-particle and bound-only SO masses are similar.

At this point, we pause to consider the meaning of our comparison at a fixed, bound-only SO mass. After infall, subhalos lose mass whereas host halos of the same initial mass keep growing. As a result, subhalos shift left in Figure~\ref{fig:frac}. This is a sensible outcome in terms of halo mass but may not reflect the evolution of galaxies, which are thought to retain (and perhaps even slightly grow) their stellar mass for some time after infall. To mimic a selection at fixed galaxy mass, we now consider the peak mass of each halo (middle column in Figure~\ref{fig:frac}). The peak is typically attained shortly before infall for subhalos \citep{behroozi_14} and commonly used in studies of the galaxy--halo connection \citep[e.g.,][]{guo_10, reddick_13}. As expected, the subhalo fraction at fixed $M_{\rm peak}$ increases compared to $M_{\rm bnd}$ because subhalos that have lost mass are now shifted into the higher-mass bin they had once attained. All definitions shift more or less in unison; the amplitude of the ratio to $\fsub$ in $R_{\rm vir,bnd}$ is slightly reduced for all definitions, masses, and redshifts.

Finally, in the right column of Figure~\ref{fig:frac}, we compare the subhalo fractions at fixed $\vmax$, another quantity that is commonly used to link halos to galaxies \citep[e.g.,][]{behroozi_19}. $\vmax$ contains unique information because it measures the potential within a radius much smaller than $\rtom$, where subhalos can more easily shield their mass from tidal disruption. The subhalo fractions at fixed $\vmax$ are similar to those at fixed $M_{\rm 200m,peak}$.

In summary, the subhalo fraction depends dramatically on the radius definition, highlighting that commonly used choices such as the ``virial'' radius are by no means unique. The subhalo fraction also depends on whether we compare halos at fixed bound mass, peak mass, or $\vmax$. At $z = 0$, $\fsub$ ranges from 6\% to 45\% at the low-mass end depending on the radius definition and halo selection, but this range would change if we could probe smaller halo masses.

\subsection{Flyby Fraction}
\label{sec:results:flyby}

Before we measure the flyby fraction, $\ffb$, we should contemplate the definition of a flyby halo and how we expect it to be affected by the halo boundary definition. We define $\ffb$ as the fraction of all host halos that were a subhalo at any time in the past. This set of halos will be composed of two distinct sub-populations: halos that had a close encounter with another, larger halo but genuinely escaped from its sphere of influence, and subhalos whose orbits have temporarily taken them outside the host halo radius. The former population should account for a small fraction of all halos and should increase with increasing halo radius (because the smaller halo is more likely to enter inside the larger halo's radius). The second population, often called ``backsplash'' halos, are orbiting their host and will eventually fall into it. Figure~\ref{fig:trees} visually demonstrates that we expect a large fraction of all subhalos to experience this type of spurious flyby event at some point if the host halo radius is small compared to the splashback radius. We expect that this population will shrink as the halo radii get larger because they will include more and more of the subhalo orbits. Given the opposite trends of genuine and spurious flyby events, the evolution of the flyby fraction with radius definition will tell us which population dominates.

At face value, the flyby fraction is easy to define: the fraction of all host halos that were a subhalo at any point along their main branch progenitor history (according to a given radius definition). In practice, applying this definition to merger trees created by almost any halo finder leads to erratic results due to spurious, temporary subhalo periods. In major mergers, for example, the host--subhalo relation can be ambiguous and switch between two halos, after which point the eventual host halo would be classified as a flyby halo. We follow the strategy of \citet{mansfield_20_ab} to eliminate such cases: we do not count a halo as a flyby if its former host is no longer alive, if the former host is a subhalo of the halo in question, or if the former host's mass is now smaller than that of the halo in question (all defined given the same radius and mass definition). In some cases, particularly for subhalo epochs at high redshift, the host may not be part of the merger trees because it never exceeded the necessary mass threshold. We also discard such events because the peak host mass was clearly smaller than the current mass of the halo in question. We emphasize that this definition of what constitutes a flyby halo is not unique. For example, we could consider future epochs to establish whether a flyby halo will eventually fall into its former host. Similarly, omitting any one of our exclusion criteria causes noticeable changes in the flyby fraction, raising the suspicion that $\ffb$ is not a particularly well-defined quantity. Nevertheless, the relative differences in flyby fractions according to different radius definitions do remain similar, which is the focus of our work. 

Figure~\ref{fig:bs} shows the flyby fraction in the \wmap cosmology at $z \approx 0$ and $z = 2$; the meaning of the lines is very similar to Figure~\ref{fig:frac} (see the Appendix for details). Regardless of the radius definition, mass variable, or redshift, $\ffb$ asymptotes to zero at the highest masses and increases toward low masses. At the smallest masses we can probe, $\ffb$ is still increasing so that we cannot put an upper bound on it. At fixed $M_{\rm 200m,bnd}$, $\ffb$ varies between about 7\% and 12\% at the low-mass end. While this range sounds relatively modest, the relative fractions differ substantially between mass definitions, especially at intermediate masses. For instance, at $M \approx 10^{12}\msunh$, using $R_{\rm sp,90\%}$ leads to only 15\% of the flyby halos found when using $\rfoc$. When plotted as a function of peak mass, the fractions are shifted to higher values (middle column of Figure~\ref{fig:bs}). In reverse, the shift means that flyby halos are more likely to have a high ratio of peak to current mass, meaning that they have lost mass at some point along their trajectory, which makes sense given that they had an encounter with a larger halo. Finally, the right column of Figure~\ref{fig:bs} shows the same results as a function of $\vmax$. In all cases, the relative differences between the radius definitions are similar. At higher redshift (bottom row), all curves are shifted to lower masses. 

Given that larger halo radii significantly reduce the flyby fraction, we conclude that the majority of flyby halos are, indeed, ``backsplash'' halos that should be classified as subhalos \citep[in agreement with][]{mansfield_20_ab}. By definition, $R_{\rm sp,90\%}$ should include at least 90\% of all subhalo orbits; in practice, it includes an even higher fraction because subhalos suffer from dynamical friction which shrinks their orbits \citep{chandrasekhar_43, adhikari_16}. In summary, splashback definitions produce significantly reduced numbers of flyby halos, which is a desirable feature (as discussed in Sections~\ref{sec:intro} and \ref{sec:discussion}).

\subsection{Are the Subhalo and Flyby Fractions Universal?}
\label{sec:results:universality}

\def\figsize{0.72}
\begin{figure}
\centering
\includegraphics[trim =  0mm 23mm 1mm 0mm, clip, scale=\figsize]{\figdir/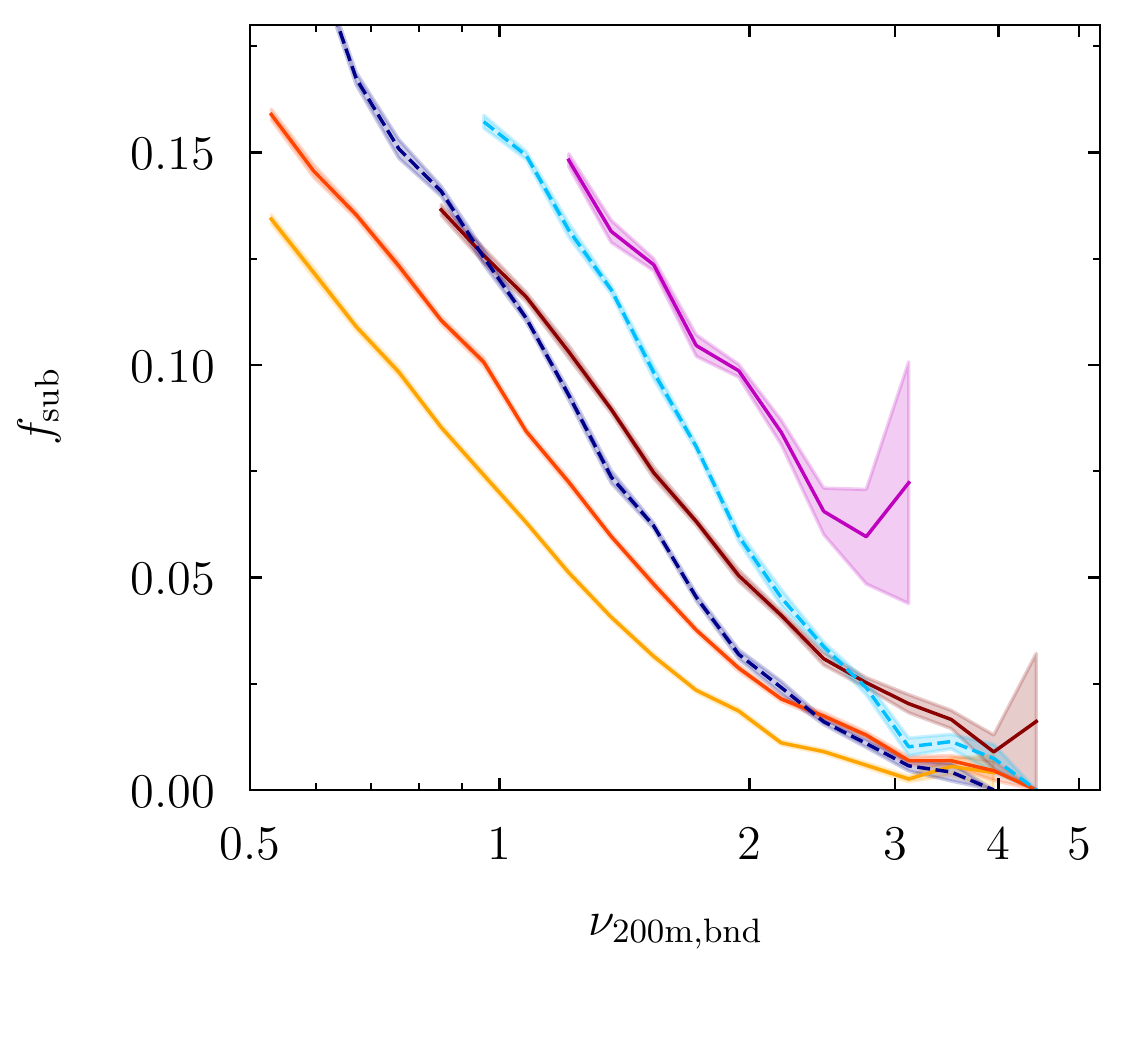}
\includegraphics[trim =  0mm 7mm 1mm 2mm, clip, scale=\figsize]{\figdir/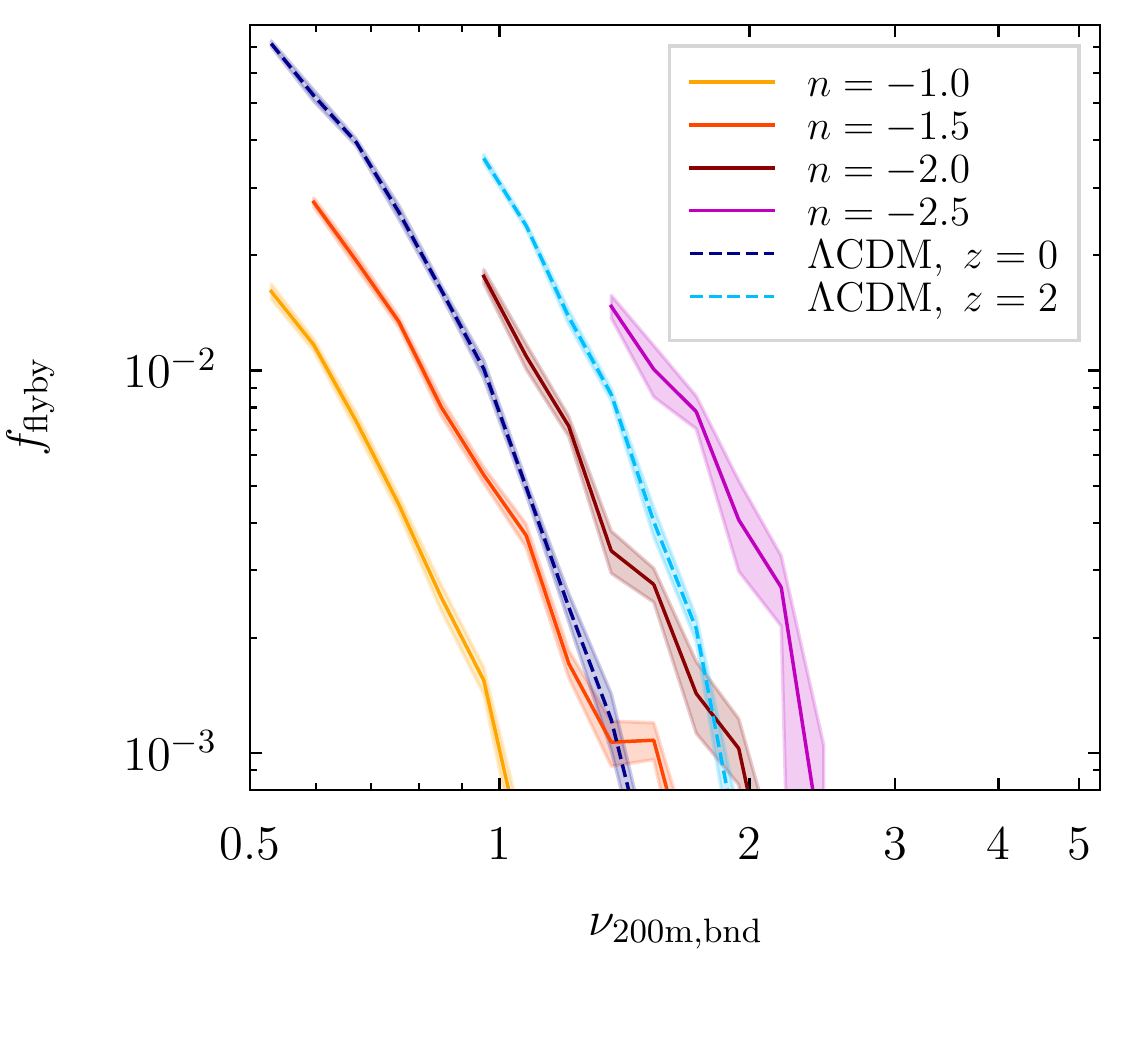}
\caption{Comparison of subhalo (top) and flyby (bottom) fractions in \LCDM and self-similar universes (according to the $R_{\rm 200m,bnd}$ definition). For comparability, the fractions are shown as a function of peak height. At fixed peak height, both fractions vary significantly with the slope of the power spectrum, $n$, in self-similar universes. The fractions in a \LCDM universe (blue dashed lines) follow qualitatively similar shapes once the slope of the power spectrum is taken into account. The flyby fraction is shown on a logarithmic scale as it reaches small values. See Section~\ref{sec:results:universality} for details.}
\label{fig:univ}
\end{figure}

\def\figsize{0.61}
\begin{figure*}
\centering
\includegraphics[trim =  0mm 7mm 1mm 2mm, clip, scale=\figsize]{\figdir/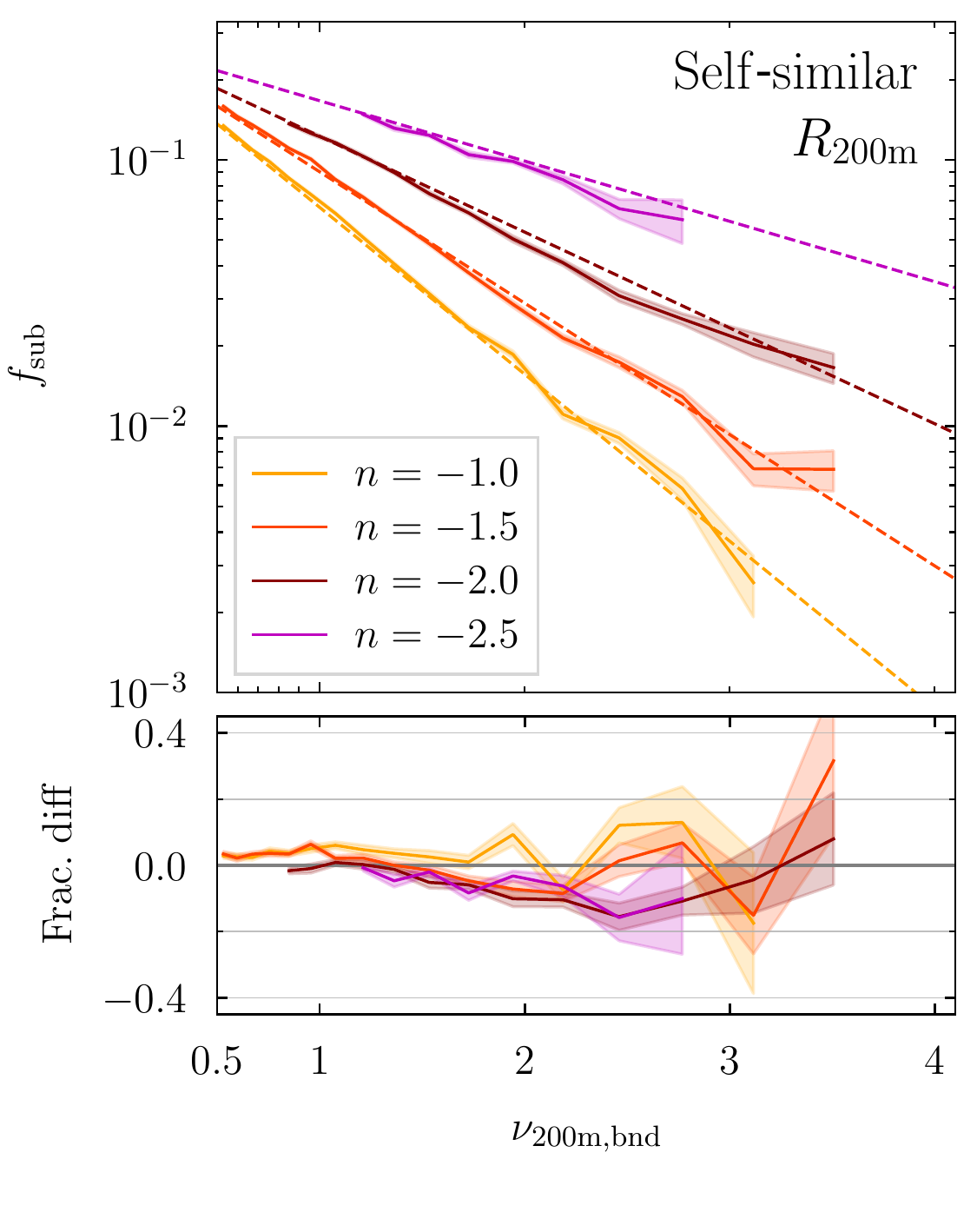}
\includegraphics[trim =  25mm 7mm 1mm 2mm, clip, scale=\figsize]{\figdir/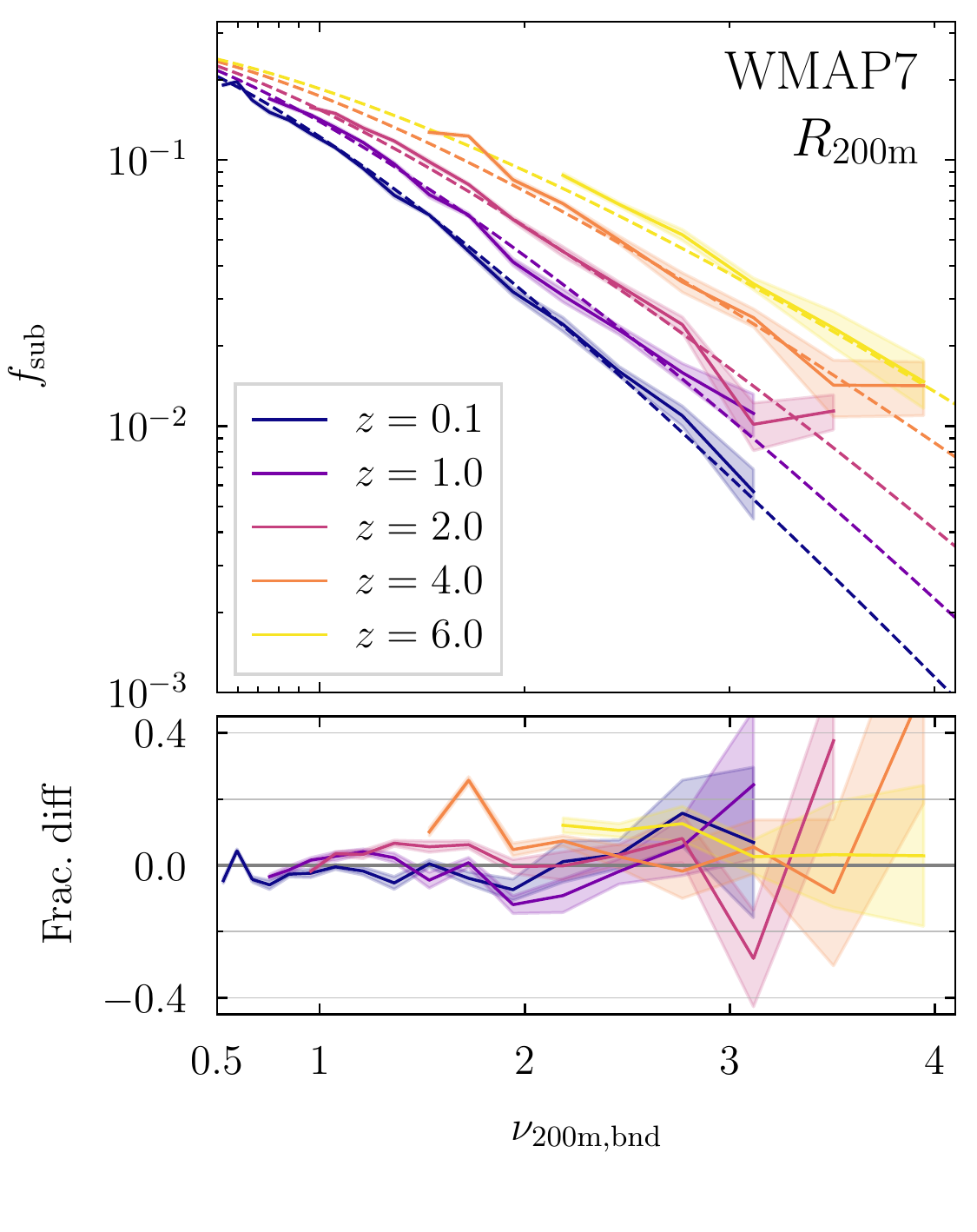}
\includegraphics[trim =  25mm 7mm 1mm 2mm, clip, scale=\figsize]{\figdir/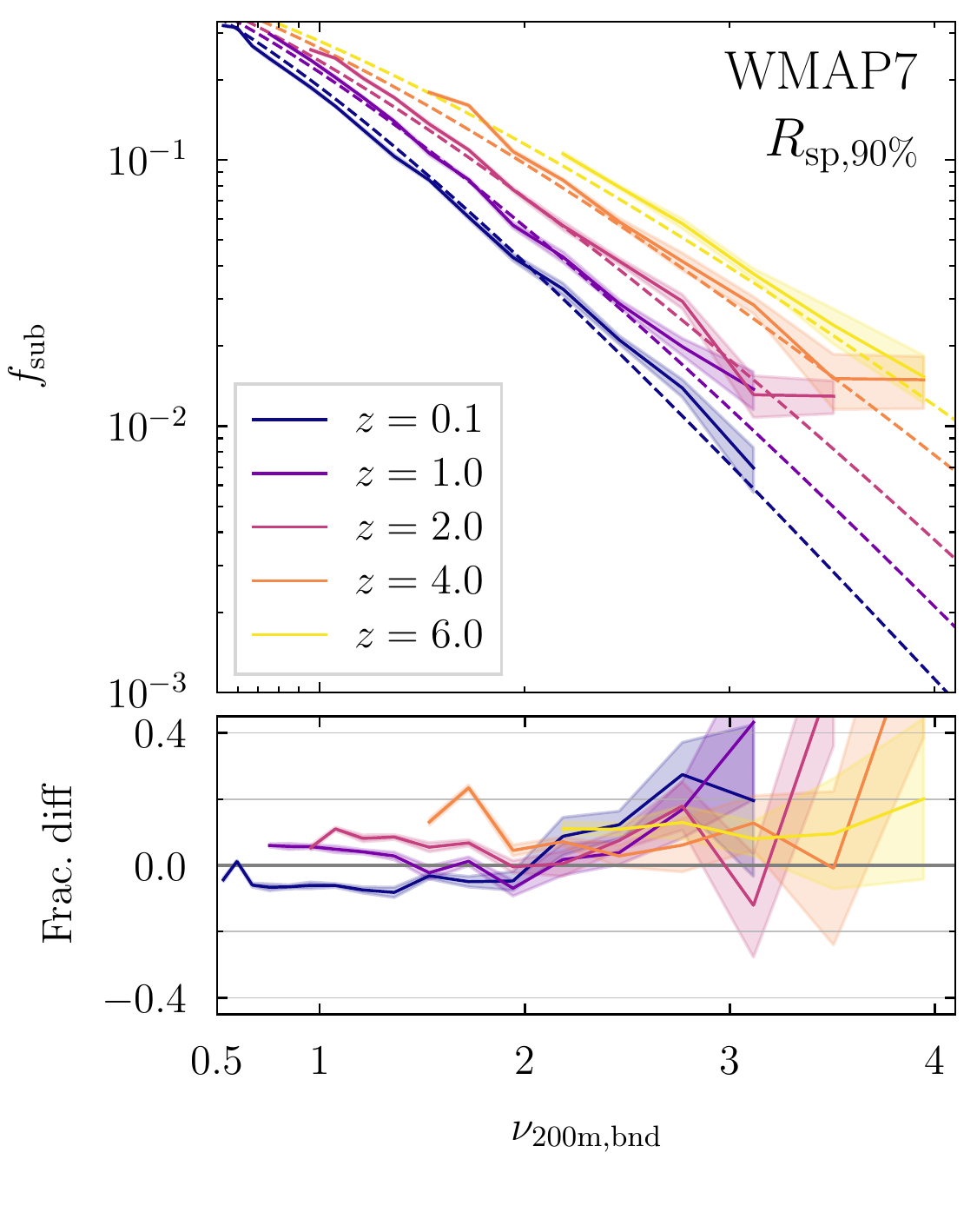}
\caption{The subhalo fraction can be understood as a function of only peak height and the linear power spectrum. In contrast to the previous figure, we show peak height on a linear and $\fsub$ on a logarithmic scale to highlight the approximately exponential nature of the relation. The bottom panels show the relative difference between the simulation data (solid lines) and the fitting function (dashed lines). In the left panel, we show $\fsub$ in the self-similar universes when using $\rtom$. In this case, the fitting function becomes an exact exponential because the slope of the power spectrum, $n$, is a constant for each simulation. When applying our fitting function to different redshifts in \LCDM (center panel), the effective slope varies, leading to deviations from an exponential. We obtain a similar fit quality when using splashback radii of any percentile (right panel). In all cases, the fitting function is accurate to better than about 20\% given the statistical uncertainties. For clarity, we have omitted bins where the statistical uncertainty is greater than a factor of 2.}
\label{fig:fit}
\end{figure*}

We now try to understand the trends in the subhalo and flyby fractions with mass, redshift, and cosmology in a unified manner. So far, we have compared $\fsub$ and $\ffb$ at fixed mass or $\vmax$, which does not allow for a fair comparison between because halo masses grow at different overall rates in different cosmologies. To facilitate such a comparison, we now express masses as peak height (as defined in Section~\ref{sec:method:defs}).

Figure~\ref{fig:univ} shows the subhalo and flyby fraction as a function of peak height. We compute the fractions and $\nu$ based on the $M_{\rm 200m,bnd}$ definition; the other SO definitions lead to qualitatively different comparisons as we discuss in Section~\ref{sec:results:fit}. The dashed blue lines show the \wmap cosmology at $z = 0$ and $z = 2$, corresponding to the orange lines in the left column of Figure~\ref{fig:frac}; the \planck cosmology gives almost identical results. The redshifts are clearly offset even in $\nu$ space: we observe both more subhalos and flybys at higher redshift. This clear trend means that $\fsub$ and $\ffb$ are not universal as a function of only peak height, i.e., that we cannot understand host--subhalo relations purely as a function of the significance of density peaks. This finding stands in contrast to, say, the halo mass function, which is approximately independent of redshift at fixed $\nu$ \citep{diemer_20_mfunc}. 

The explanation for the non-universality is provided by the self-similar simulations, which are distinguished by different slopes of the power spectrum, $n = d\ln P / d \ln k$ (solid lines in Figure~\ref{fig:univ}). In these universes, all redshifts give the same results at fixed $\nu$ and have been combined into one curve per simulation (Appendix~\ref{sec:app}). Clearly, shallower $n$ lead to lower subhalo and flyby fractions at fixed peak height. This trend may be counterintuitive because a shallower power spectrum means that, at a given scale, there is more substructure to be accreted into halos. However, this logic is reversed here: at fixed peak height, there are more, larger, potential host halos in cosmologies with a steeper power spectrum slope, leading to a higher subhalo fraction.

\begin{deluxetable}{lccc}
\tablecaption{Best-fit parameters for the subhalo fraction model
\label{table:fits}}
\tablewidth{0.47\textwidth}
\tablehead{
\colhead{Parameter} &
\colhead{$R_{\rm 200m}$} &
\colhead{$R_{\rm sp,mn}$} &
\colhead{$R_{\rm sp,\%}$}
}
\startdata
\rule{3pt}{0pt} $a_0$           & $   1.270$ & $   0.965$ & $   1.439$ \\
\rule{3pt}{0pt} b               & $   2.057$ & $   2.069$ & $   2.042$ \\
\rule{3pt}{0pt} c               & $   0.614$ & $   0.550$ & $   0.537$ \\
\rule{3pt}{0pt} $\kappa$        & $   1.574$ & $   2.032$ & $   2.106$ \\
\rule{3pt}{0pt} $a_{\rm p}$     &         $0$ &         $0$ & $  -0.865$
\enddata
\end{deluxetable}

We can now try to equate the redshift trend in \LCDM to a trend in the power spectrum slope, which we measure at the Lagrangian scale of halos (Equation~\ref{eq:rl}). This slope varies between about $-2$ for the largest halos at $z = 0$ and $-3$ for small halos at high redshift \citep[e.g.,][]{diemer_15}. The trends in the \LCDM and self-similar universes seem generally compatible, as we find more subhalos at steeper slopes and higher redshifts in \LCDM. On the other hand, Figure~\ref{fig:univ} suggests that the slopes might not match: we would expect the $z = 0$ lines to lie between $n = -2$ and $-2.5$ while they are closer to $-1.5$ at high peak heights. However, given that a halo can be a subhalo of any halo that is larger than itself, the subhalo and flyby fractions are likely influenced by the shape of $P(k)$ over a wide range of $k$ scales rather than at only the Lagrangian scale. We try to crudely capture this larger range by defining an effective slope,
\begin{equation}
\label{eq:neff}
\neff(M) = -2 \left. \frac{d\ln \sigma(R)}{d \ln R} \right \vert_{R = \kappa R_{\rm L}}-3 \,.
\end{equation}
For the power-law power spectra of the self-similar simulations, $\neff = n$. In \LCDM, $\neff$ depends on $P(k)$ over all scales that significantly contribute to the variance $\sigma(R)$. We measure the slope of $\sigma$ near the Lagrangian radius but allow a scaling via the free parameter $\kappa$. Even setting $\kappa = 1$, we find that $\neff$ in the \wmap cosmology is similar to the values of $n$ where the lines of $\fsub$ overlap in Figure~\ref{fig:univ}. The effective slope can easily be evaluated using the \colossus code; it is discussed in detail in \citet{diemer_19_cm}.

\subsection{Fitting Function for the Subhalo Fraction}
\label{sec:results:fit}

We now construct a universal fitting function for $\fsub$ with $\nu$ and $\neff$ as input variables. We note that the peak height trend of $\fsub$ in the self-similar cosmologies (left panel of Figure~\ref{fig:fit}) is well fit by an exponential,
\begin{equation}
\label{eq:fit}
\fsub = \exp\left( -\left[ a + \left(b + c \times \neff \right) \times \nutom \right] \right) \,,
\end{equation}
where $a$ is a normalization, $b$ controls how $\fsub$ scales with peak height, and $c$ introduces an additional, linear dependence on the effective slope of the power spectrum. Remarkably, the same function fits both the self-similar and \LCDM cosmologies (center panel of Figure~\ref{fig:fit}); to achieve this match, we let $\kappa$ be a fourth free parameter. The fact that $\kappa > 1$ means that $\fsub$ is most sensitive to the slope of $\sigma$ corresponding to larger halos, i.e., the hosts of subhalos at the given peak height. When only fitting \LCDM, $\kappa$ is largely degenerate with the normalization $a$. The self-similar simulations break this degeneracy because they are insensitive to $\kappa$ because $\neff = n$ is a constant.

So far, we have considered only $\rtom$ as a definition of the halo radius. Interestingly, the virial and critical-density SO definitions $\rvir$, $\rtoc$, and $\rfoc$ lead to irreconcilable differences between \LCDM and the self-similar universes. They cannot be fit with Equation~\ref{eq:fit}, but it is clear that this is not an issue with the fitting function: the scaling of $\fsub$ with $\nu$ and $\neff$ is simply not compatible with the self-similar simulations. For $\rtoc$, this result is intuitive because it is exactly equivalent to $\rtom$ at all redshifts in Einstein--de Sitter cosmologies. In \LCDM, their evolution diverges strongly at low redshift. Thus, a function of only $\nu$ and $\neff$ cannot fit $\fsub$ based on both $\rtom$ and $\rtoc$.

Finally, we apply our fitting function to the splashback definitions. There is no physically meaningful way to define the current splashback mass of a subhalo. Thus, we retain $\nutom$ as the input variable but measure $\fsub$ based on the respective splashback radii (right panel of Figure~\ref{fig:fit}). We find that we can accommodate the different percentiles with a simple, linear scaling of the normalization, where $a = a_0$ for $\rtom$ and $R_{\rm sp,mn}$ and $a = a_0 + a_{\rm p} \times p$ for the percentile definitions. Here, $p$ is the percentile divided by $100$, e.g., $0.75$ for the 75th percentile. 

We constrain the best-fit parameters using a Levenberg--Marquart least-squares minimization (Table~\ref{table:fits}). We simultaneously fit to all data from the self-similar, \wmap, and \planck cosmologies; for the latter two, we include redshifts, $0.13$, $0.5$, $1$, $2$, $4$, and $6$. We add a systematic uncertainty of 3\% in quadrature to the statistical uncertainties to prevent the most statistically significant bins from dominating the fit. We separately fit for three sets of parameters for $\rtom$, $R_{\rm sp,mn}$, and all percentile-based definitions of $\rsp$. However, the $R_{\rm sp,mn}$ fit is very similar to $R_{\rm sp,50\%}$. As discussed above, the other SO definitions cannot be reasonably fit with a function of only $\nu$ and $\neff$. We obtain $\chi^2$ values per degree of freedom between $1.8$ and $2.8$, but these values are meaningless as they depend on the added systematic error. The bottom panels of Figure~\ref{fig:fit} demonstrate that our fitting function is accurate to about 20\% (taking into account the statistical uncertainties). The accuracy is similar for all mass definitions and cosmologies.

However, the main point of our fitting function is not accuracy but rather to demonstrate that such a function exists. The simplicity of our four-parameter fit is remarkable, given the complexity of the host--subhalo calculations. If accuracy was the goal, we could slightly improve it by fitting only \LCDM (without the self-similar simulations) and by fitting the splashback percentiles separately. It is also remarkable that a universal function exists for $\rtom$ but not for $\rtoc$, suggesting that SO definitions based on the mean density lead to more physically meaningful subhalo assignments. Either way, it is reassuring that all splashback radii produce equally universal results. We refrain from constructing a similar fitting function for $\ffb$ because its values depend strongly on the exact criteria for flybys (Section~\ref{sec:results:flyby}).


\section{Discussion}
\label{sec:discussion}

\begin{figure*}
\centering
\includegraphics[trim =  0mm 0mm 0mm 0mm, clip, width=\textwidth]{\figdir/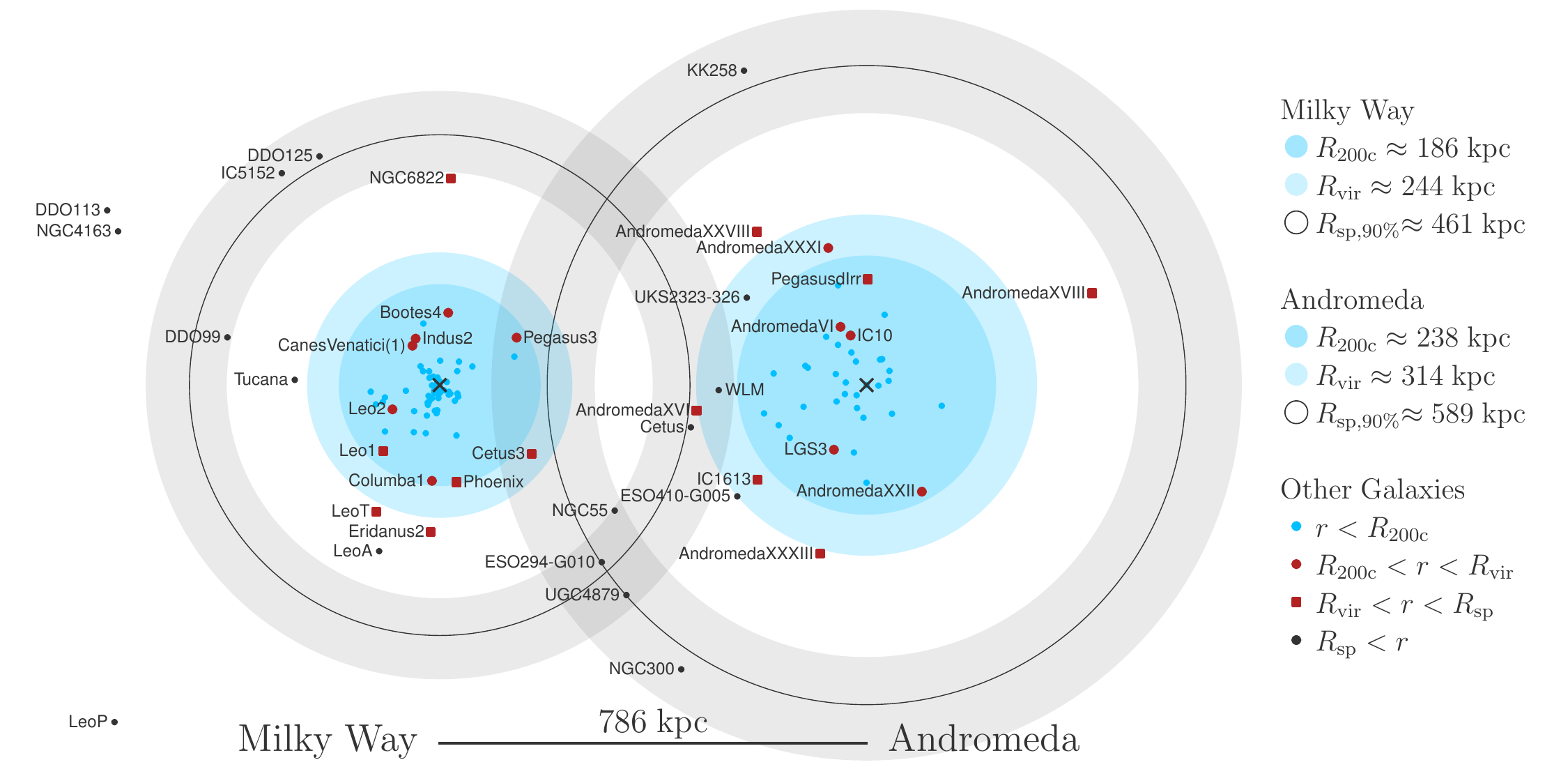}
\caption{Halo radii in the Local Group and the resulting classification of satellites and field galaxies. The frame is rotated such that the Milky Way--Andromeda axis is in the $x$ direction. The blue shaded areas highlight the commonly used $\rtoc$ and $\rvir$ boundaries; light-blue dots mark the positions of satellites inside $\rtoc$. Red dots and squares indicate galaxies that are considered satellites if we switch to $\rvir$ and $\rsp$ (black rings, taken to be the radius that includes 90\% of apocenters). Numerous objects that have conventionally been considered field or flyby (``backsplash'') galaxies are likely inside the splashback radius and are thus better thought of as satellites. The virial masses were taken from \citet{diaz_14} and converted to $\rtoc$ using the concentration--mass relation of \citet{diemer_19_cm}. The splashback radius was estimated using the formula of \citetalias{diemer_20_catalogs}; the shaded area shows the $1\sigma$ uncertainty of 0.07 dex. The galaxy positions are from the updated catalogs of \citet{mcconnachie_12}.}
\label{fig:local}
\end{figure*}

We have demonstrated that the definition of the halo boundary has a dramatic impact on the distinction between hosts and subhalos. In this section, we discuss potential consequences for our understanding of the Local Group (Section~\ref{sec:discussion:local}) and for models of galaxies and large-scale structure (Section~\ref{sec:discussion:halomodel}). We end by highlighting a number of intriguing questions for future work (Section~\ref{sec:discussion:future}).

\subsection{A Splashback Perspective of the Local Group}
\label{sec:discussion:local}

The classification of halos as isolated, subhalos, and flybys is particularly important in the Local Group (LG). First, the LG can be used as a laboratory for the interaction of dwarf galaxies with a dense environment. For example, the local dwarfs are known to be gas poor and slowly star-forming in comparison to field galaxies \citep[e.g.,][]{grcevich_09, geha_12, spekkens_14}. To make such a comparison, however, we need to categorize the dwarfs as satellites and flybys in a meaningful way \citep[e.g.,][]{simpson_18}. Second, the orbits of satellites are used to infer the masses of the Milky Way (MW) and Andromeda (M31), but these calculations often depend on whether a satellite is on its first infall or not \citep[e.g.,][]{boylankolchin_13, blana_20}. Similarly, measurements of the combined LG mass via dynamics rely on correct modeling of the mass distribution outside $\rvir$ \citep{penarrubia_17}. Third, many of the satellites in the LG are known to lie in planar, coherently rotating structures \citep{ibata_13, shaya_13, pawlowski_14, libeskind_15}. 

Motivated by these applications, there has been a lively discussion as to whether certain dwarfs in the LG are field galaxies, orbiting the MW or M31, or flyby (``backsplash'') galaxies \citep[e.g.,][]{besla_07, teyssier_12, pawlowski_14, buck_19, blana_20, mcconnachie_20_solo}. This distinction necessarily depends on a halo radius, which is generally assumed to be $\rvir$ or $\rtoc$. Our results demonstrate that these definitions lead to artificial classifications of orbiting satellites as flyby and field galaxies even though they reside within the splashback radius. The question of whether a galaxy is on a first infall or orbiting is physically meaningful, but it should be judged according to whether the galaxy has undergone a pericenter rather than whether it has passed through an arbitrary radius such as $\rtoc$. Moreover, classifying dwarfs as flybys seems to imply that they will escape from the LG eventually, but our results show that the vast majority of galaxies within $\rsp$ do eventually join the host halo.

Figure~\ref{fig:local} visualizes how our picture of the LG changes when we reframe it in terms of splashback radii \citep[see also][]{buck_19, deason_20_galaxies}. We compare the SO radii $\rtoc$ and $\rvir$ to $R_{\rm sp,90\%}$, chosen to include virtually all local satellite orbits. For halos with MW and M31 masses, we find a median $R_{\rm sp,90\%} \approx 2.5 \rtoc$. However, we emphasize that these splashback radii are merely a guess based on the fitting function of \citetalias{diemer_20_catalogs}; they have not been measured yet and we do not know the accretion rate of the LG halos. The gray rings indicate the resulting 0.07 dex $1\sigma$ uncertainty \citepalias{diemer_20_catalogs}. Regardless of whether the LG splashback radii are below or above the median, our picture of the LG changes fundamentally. Instead of two halos that are close but separate, we should think of the MW and M31 halos as partially overlapping, which can also be seen in visualizations of simulated LG analogs \citep{libeskind_10, garrisonkimmel_14_elvis}.

Naturally, switching to $\rsp$ also affects our understanding of LG satellites. The red points in Figure~\ref{fig:local} mark the positions of galaxies that would be classified as satellites within $\rsp$ but not within $\rtoc$. The strong selection effect against distant dwarfs means that we will find many more galaxies in this radial range. A large fraction of this population would be classified as flybys if $\rtoc$ is used \citep{simpson_18, blana_20}; in the splashback picture, they are simply satellites that have had at least one pericenter. Considering the dynamics of the LG dwarfs is beyond the scope of this paper, but we can comment on their classification purely based on positions. For example, \citet{kirby_14} found that the internal dynamics of isolated LG dwarfs are not systematically different from satellites. However, some of their targets (such as Pegasus dIrr and IC 1613) probably lie within the splashback radius of M31, which could help to explain their satellite-like nature. 

In terms of the flyby fraction, \citet{teyssier_12} compared the positions and velocities of LG dwarfs to simulations and inferred a high probability for about 13\% of them to be flybys. Using a similar technique, \citet{buck_19} find lower flyby probabilities and point out that the likely flybys lie within the presumed splashback radii of MW and M31. Based on simulated orbits, \citet{mcconnachie_20_solo} are even more pessimistic about the flyby probabilities of some of these galaxies \citep[see also][]{blana_20}. We can statistically assess these results based on the flyby fractions of Figure~\ref{fig:bs}. The farther-out LG dwarfs are thought to inhabit halos with a wide range of peak masses from a few times $10^8$ to about $10^{11} \msun$ \citep{mcconnachie_12, garrisonkimmel_17, fattahi_18, buck_19}. At the smallest halo masses we can test, $M_{\rm peak} \approx 10^9 \msunh$, we infer $\ffb \approx 5\%$ when using $\rtoc$ and about 2\% when using $R_{\rm sp,90\%}$. The strong evolution in the backsplash fraction with mass makes it difficult to predict a single number, but it seems unlikely that the flyby fraction in the LG should be as high as 10\%. On the other hand, our flyby fractions refer to the total halo sample, not to halos close to larger neighbors. Regardless, we note that most of the flyby candidates of \citet{teyssier_12} lie well outside of $R_{\rm sp,90\%}$ and would thus be true flyby halos if they interacted with the LG, but NGC 6822, Phoenix, and Leo T currently reside around $0.9 R_{\rm sp,90\%}$ \citep[see also][]{blana_20}. 

Another phenomenon that is easily explained in the splashback picture are so-called ``renegade'' subhalos that switch hosts between the MW and M31 \citep{knebe_11_renegade}. While no such satellites have been reliably identified in the LG, Leo T is a candidate \citep{mcconnachie_20_solo}. Renegade subhalos would be a natural consequence of the overlapping splashback radii. All of our considerations highlight the importance of using a physical halo boundary when discerning between satellites, field galaxies, and flybys.

\vspace{1cm}

\subsection{Galaxy and Halo Modeling}
\label{sec:discussion:halomodel}

Observationally, the issue of where to draw the halo boundary is perhaps most apparent at the transition between the collapsed matter inside halos and the large-scale structure around them \citep[the so-called ``1-halo'' and ``2-halo'' terms; e.g.,][]{hayashi_08}. This transition manifests itself as a break in the overall density profile and as a resulting dip in the lensing signal \citep[e.g.,][]{leauthaud_11, oguri_11, tully_15, tomooka_20}. If this region is interpreted based on too small a halo boundary, one might, for example, conclude that some of the 2-halo signal is due to flyby halos \citep{sunayama_16}. Another observable that might be impacted by the host--subhalo (or central--satellite) assignment is the total stellar mass within a halo \citep[e.g.,][]{lin_04_icl, gonzalez_07, leauthaud_12_total}. This statistic has recently received renewed attention due to its tight connection to halo mass \citep{tinker_20_total, bradshaw_20, demaio_20, huang_20_hsc}. The scatter in this relation should be smallest if all satellites within a group or cluster are considered, not some subset inside a smaller halo boundary.

Theoretically, the 1-halo and 2-halo clustering regimes are generally understood based on the so-called ``halo model,'' which posits that all matter resides in halos \citep[e.g.,][]{ma_00_halomodel, seljak_00, zentner_05}. On small scales, the clustering follows the halo density profile; on large scales, it follows the linear correlation function times some halo bias \citep{cole_89}. The definition of the halo boundary thus matters for the halo model's predictions and its interpretation. This interplay was recently investigated by \citet{garcia_20}, who left the halo radius as a free parameter and found that a large radius, possibly larger than $R_{\rm sp,90\%}$, provides the best fit to the clustering in simulations. This intriguing result hints at the possibility of constructing a halo model based on the splashback radius. 

When adding galaxies to our modeling of large-scale structure, adopting a splashback boundary could affect the results via the host--subhalo distinction but also via systematic changes in host masses, for example, due to environment-dependent mass accretion rates. Given the subject of this paper, we focus on the former effect. We consider three popular techniques to infer the galaxy--halo connection \citep{wechsler_18}. First, subhalo abundance matching (SHAM) assigns galaxies to halos by matching rank-ordered lists of a galaxy property, such as stellar mass, and a halo property, such as halo mass \citep[e.g.][]{kravtsov_04_hod, vale_04, conroy_06}. While the SHAM assignment does not necessarily distinguish between hosts and subhalos, the results are sometimes validated against observed satellite fractions based on a group finder \citep[e.g.,][]{yang_05_groupfinder, tinker_11, reddick_13, lehmann_17}. However, the differences in $\fsub$ due to the halo boundary can be larger than those due to the physics included in the SHAM model, such as the halo property ($M_{\rm peak}$, $\vmax$ etc.) or the scatter in the stellar mass--halo mass relation \citep{behroozi_10, reddick_13}. Thus, the conclusions drawn from comparisons to observed satellite fractions might change depending on the definition of the halo boundary. Second, in a halo occupation distribution (HOD) analysis, we assign one central and any number of satellite galaxies to a halo based on its mass \citep[e.g.,][]{peacock_00, seljak_00, berlind_02, cooray_02} or other halo properties \citep[e.g.,][]{hearin_16}. This assignment is not sensitive to changes in the abundance of subhalos in the halo catalogs, but a larger halo boundary would mean removing hosts that are, by construction, strongly clustered around other halos. The free parameters of the HOD would readjust to match the observations (e.g., clustering signals), possibly leading to a different physical interpretation of the results. Third, semi-analytical models (SAMs) constitute simplified descriptions of the sophisticated processes of galaxy formation that are applied to simulated merger trees \citep{kauffmann_93, cole_94_sam, somerville_99, benson_12, croton_16, lagos_18_shark}. The impact of the host--subhalo assignment will depend on whether a specific model treats subhalos differently from hosts, for example, by explicitly modeling satellite stripping and disruption \citep[e.g.,][]{guo_11, stevens_16_darksage}. In summary, we expect the halo boundary definition to have some impact on most types of galaxy--halo modeling, both due to changed host masses and due to the host--subhalo assignment; the importance of these effects will need to be quantified model by model.

\subsection{Future Directions}
\label{sec:discussion:future}

We have left a number of theoretical and numerical issues for future investigations. For example, \citet{villareal_17} showed that assembly bias \citep{gao_05_ab} can be mitigated by choosing large halo boundaries, an effect that can now be quantified for splashback radii. On the other hand, the question of assembly bias also highlights a big caveat: our results are derived from spherical halo radii, whether splashback or SO. Recently, \citet{mansfield_20_ab} showed that non-sphericity leads to significant differences in the subhalo assignment and thus in the assembly bias signal.

One somewhat unsatisfying aspect of our results is that both $\fsub$ and $\ffb$ are rising at the lowest halo masses that we can access, meaning that we cannot constrain their asymptotic values at low mass. Our fitting function suggests that $\fsub$ should approach a finite value, but that remains to be tested. It is possible that the asymptotic value of $\fsub$ may depend on the smallest possible halo mass and thus on the cutoff scale of the power spectrum (e.g. due to warm dark matter). If this cutoff allowed for, say, Earth-mass halos \citep[e.g.,][]{diemand_05}, it is conceivable that the vast majority of the smallest halos would be subhalos. Testing this hypothesis may demand simulations with an unprecedented dynamic range \citep[e.g.,][]{wang_20_zoom}. 

Finally, we have made no attempt to correct for numerical issues that lead to unphysical subhalo disruption, for example, by tracking undetectable subhalos based on a subset of their constituent particles \citep[``orphans'' or ``cores'', e.g.,][]{wang_06_orphans, heitmann_19}. We will return to this issue in Diemer \& Behroozi 2021 (in preparation), where we track so-called ``ghost'' subhalos and propose a new definition of their mass.


\section{Conclusions}
\label{sec:conclusion}

We have systematically investigated the fraction of halos that are a subhalo or flyby halo based on both spherical overdensity and splashback definitions of the halo boundary. We find that both fractions depend strongly on the radius definition. Our main conclusions are as follows.
\begin{enumerate}
\item The subhalo fraction depends on the chosen definition of subhalo mass, with lower subhalo fractions at fixed bound-only mass than at fixed peak mass or $\vmax$.
\item Compared to the commonly used $\rvir$, defining subhalos via $\rfoc$ leads to up to 60\% fewer subhalos while using splashback radii leads to between 50\% and 100\% more subhalos at the low-mass end. The differences are slightly smaller at higher redshift but generally persist across cosmic time and cosmology.
\item The flyby fraction follows the opposite trend, where larger radii lead to fewer flyby halos. This trend demonstrates that the vast majority of flyby halos are ``backsplash'' satellites that should be classified as subhalos. A subhalo assignment based on the splashback radius largely eliminates this issue.
\item The subhalo fraction can be understood as a function of only peak height and the slope of the power spectrum. We present a simple, universal fitting formula for the subhalo fractions based on $\rtom$ and splashback radii.
\item Our understanding of the LG and its dwarf galaxies changes significantly when using $\rsp$ as a the halo boundary.
\end{enumerate}
We have left numerous open questions for future work, particularly regarding the impact of the radius definition on the galaxy--halo connection. Our catalogs and merger trees are publicly available at \href{http://www.benediktdiemer.com/data}{benediktdiemer.com/data}; we hope that this paper provides motivation for further investigations.

\def\figsize{0.54}
\begin{figure*}
\centering
\includegraphics[trim =  1mm 23mm 1mm 2mm, clip, scale=\figsize]{\figdir/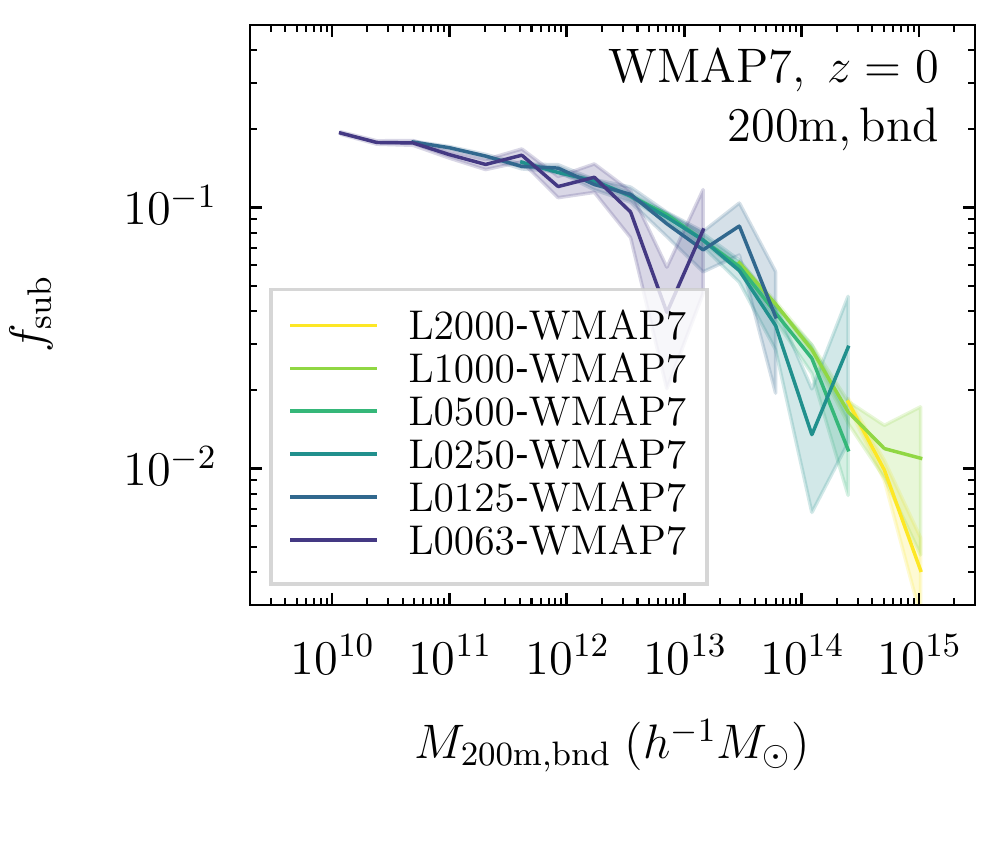}
\includegraphics[trim =  25mm 23mm 1mm 2mm, clip, scale=\figsize]{\figdir/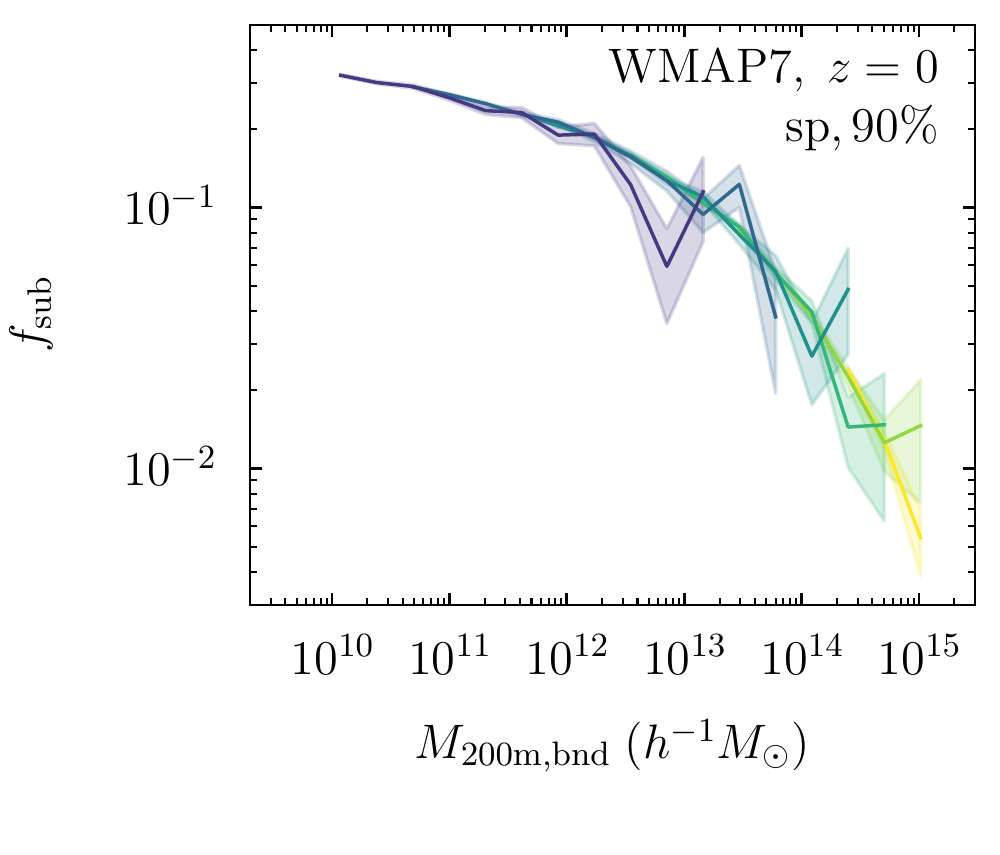}
\includegraphics[trim =  25mm 23mm 1mm 2mm, clip, scale=\figsize]{\figdir/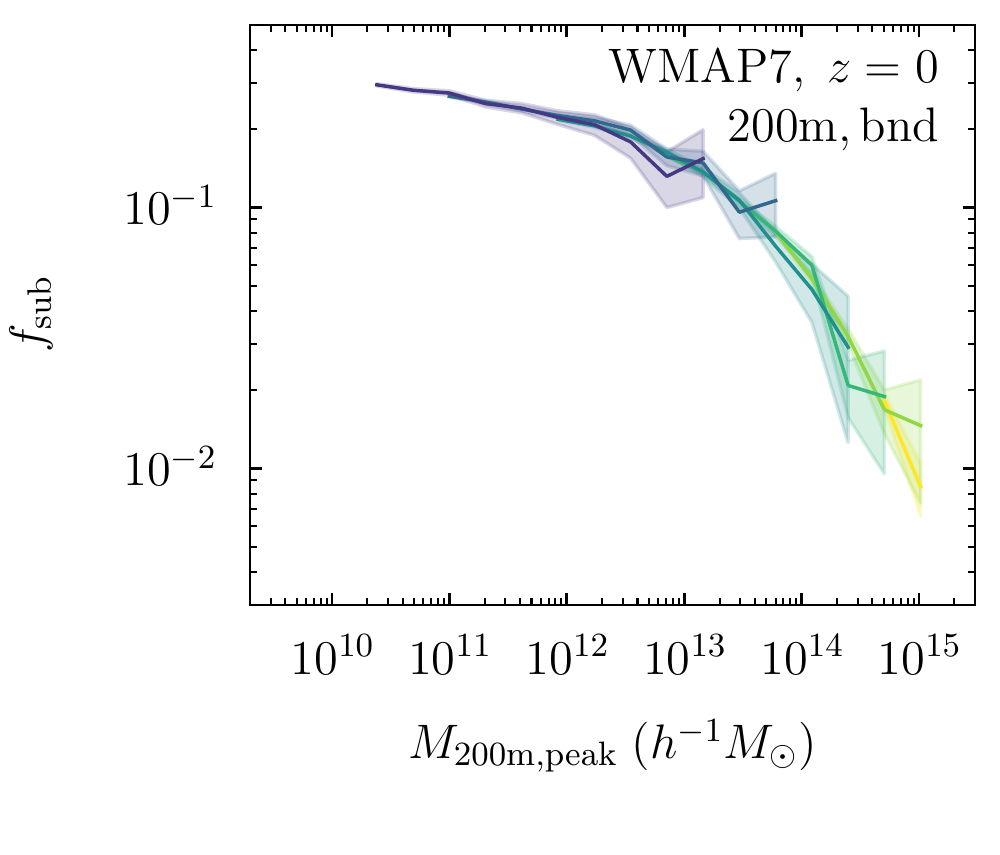}
\includegraphics[trim =  25mm 23mm 2mm 2mm, clip, scale=\figsize]{\figdir/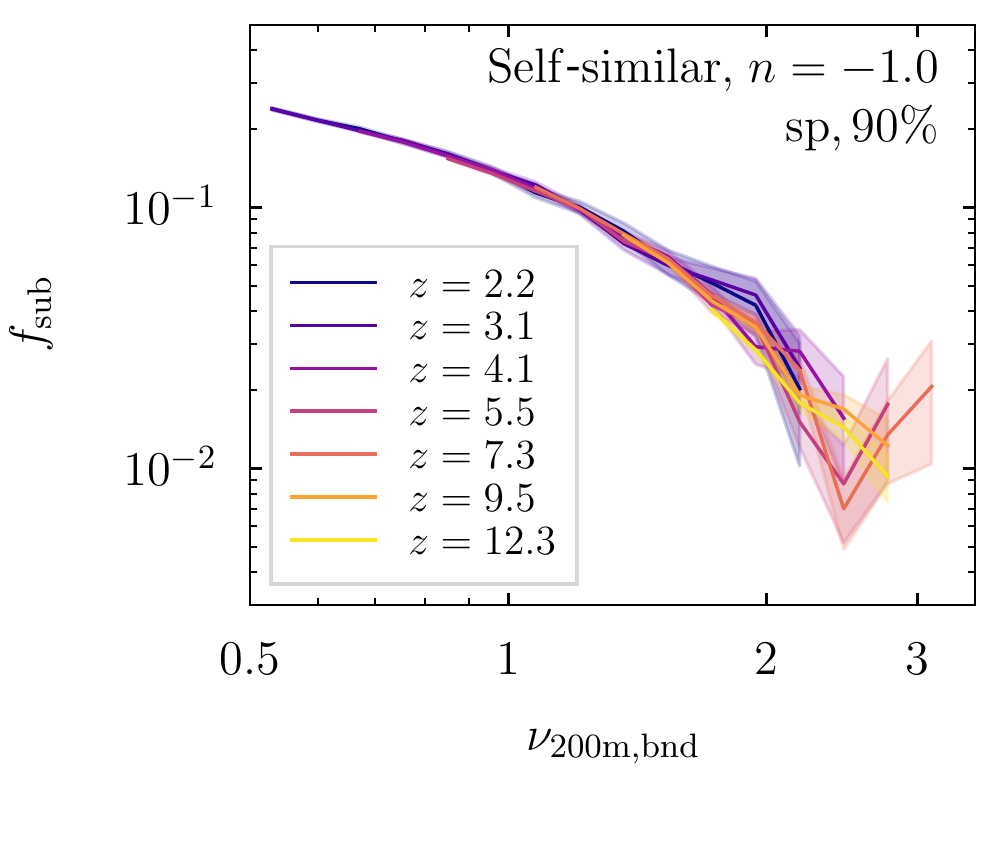}
\includegraphics[trim =  1mm 7mm 1mm 2mm, clip, scale=\figsize]{\figdir/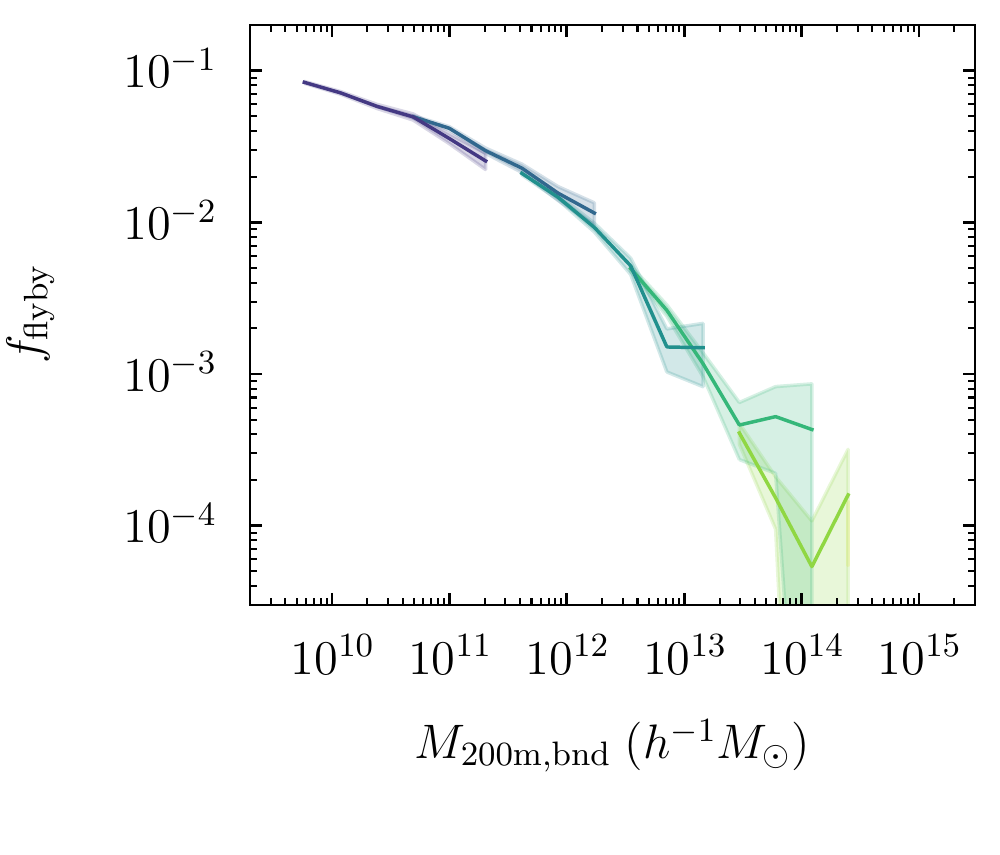}
\includegraphics[trim =  25mm 7mm 1mm 2mm, clip, scale=\figsize]{\figdir/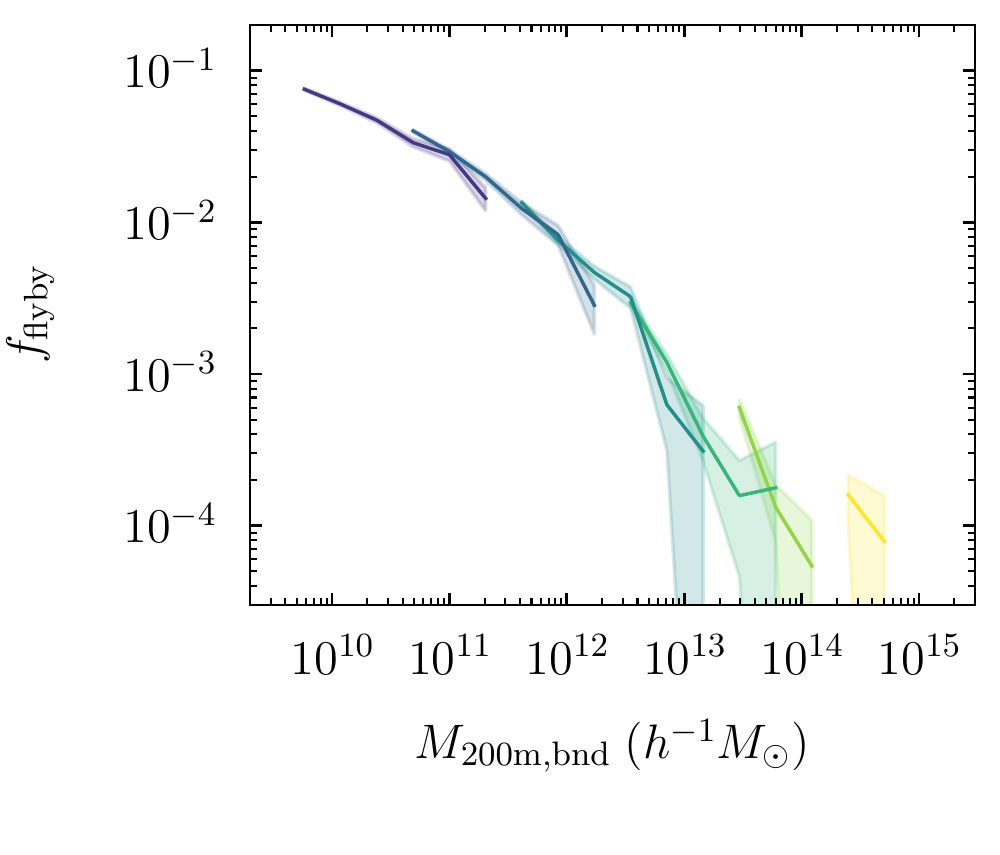}
\includegraphics[trim =  25mm 7mm 1mm 2mm, clip, scale=\figsize]{\figdir/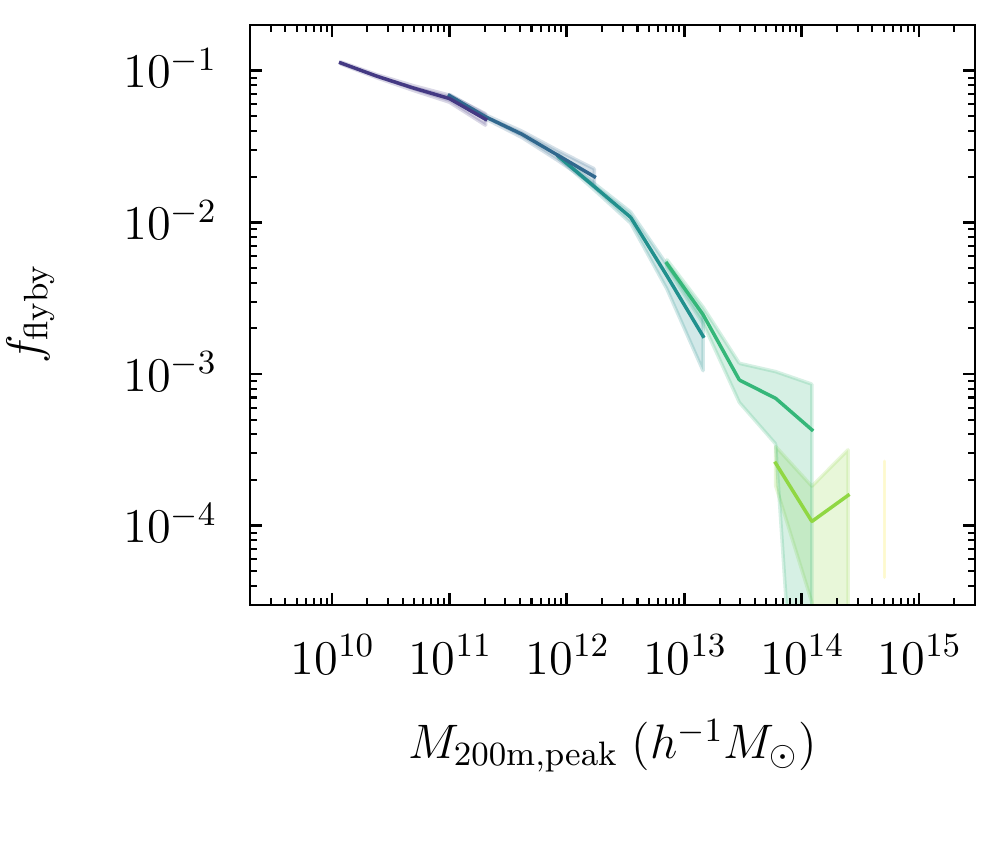}
\includegraphics[trim =  25mm 7mm 2mm 2mm, clip, scale=\figsize]{\figdir/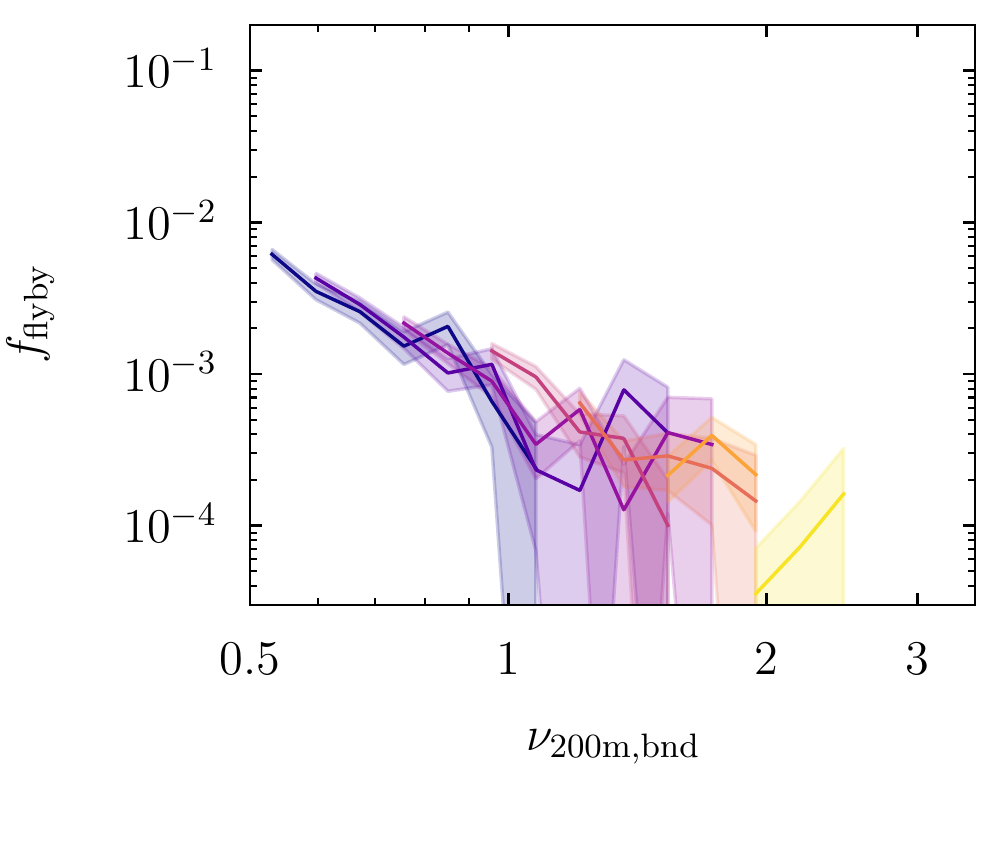}
\caption{Convergence of the subhalo (top) and flyby (bottom) fraction between different \LCDM simulations (left three columns) and different redshifts in a self-similar simulation (right column). The left two columns show results binned by the current $\mtom$ of halos, with subhalo and flyby fractions based on $\rtom$ and on $R_{\rm sp,90\%}$. The third column shows results binned by peak mass, which leads to different convergence properties. The fourth column is binned in the current peak height of halos and compares different redshifts in the $n = -1$ self-similar simulation. In the bottom row, we omit bins that contain fewer than 2000 halos as they lead to very noisy estimates given the small flyby fractions. The subhalo and flyby fractions are converged to the accuracy expected given the statistical uncertainties, although slight convergence issues are apparent in the flyby fraction in self-similar simulations.}
\label{fig:conv1}
\end{figure*}

\vspace{0.5cm}

I am grateful to Han Aung, Peter Behroozi, Mat{\'\i}as Bla{\~n}a, Joe DeRose, Michael Joyce, Philip Mansfield, Surhud More, Daisuke Nagai, and Enia Xhakaj for productive discussions and feedback on a draft. I am especially thankful to Andrew Hearin for his feedback and creative input. I thank the anonymous referee for their insightful comments, which significantly improved the paper. This work was partially completed during the coronavirus lockdown and would not have been possible without the essential workers who did not enjoy the privilege of working from the safety of their homes. All computations were run on the \textsc{Midway} computing cluster provided by the University of Chicago Research Computing Center. This research made extensive use of the Python packages \textsc{NumPy} \citep{code_numpy2}, \textsc{SciPy} \citep{code_scipy}, \textsc{Matplotlib} \citep{code_matplotlib}, and \colossus \citep{diemer_18_colossus}. This research was supported in part by the National Science Foundation under Grant No. NSF PHY-1748958. Support for Program number HST-HF2-51406.001-A was provided by NASA through a grant from the Space Telescope Science Institute, which is operated by the Association of Universities for Research in Astronomy, Incorporated, under NASA contract NAS5-26555. 


\appendix

\section{Technical Details}
\label{sec:app}

In this appendix, we describe how we compute the multi-simulation datasets of $\fsub$ and $\ffb$ that are shown in Figures~\ref{fig:frac}--\ref{fig:fit}. 

\subsection{Resolution Limits and Convergence}
\label{sec:app:convergence}

We begin by determining bins in $M_{\rm X}$, $M_{\rm peak}$, $\vmax$, or $\nu$. Halos from all simulations of a given cosmology will contribute to a given bin as long as they are resolved with a minimum number of particles in this simulation, $N_{\rm min}$. This limit will ensure that the different radii are properly measured and that the catalogs are complete \citepalias{diemer_20_catalogs}. We require that the entire mass range in a bin must be resolved with more than $N_{\rm min}$ particles; otherwise we omit the given simulation from the bin to avoid partial contributions.

However, a fixed $N_{\rm min}$ will select different halos depending on the mass definition, because $\msp \gsim \mtom \geq \mtoc > \mfoc$. To apply a comparable limit to all SO definitions, we define $N_{\rm min}$ to refer to $\mtom$. For the other SO masses, we convert the bin edges' masses assuming a Navarro-Frenk-White profile \citep{navarro_97} and the \citet{diemer_19_cm} mass--concentration relation. This procedure ensures a more or less uniform cut for all definitions. For the splashback definitions, we use the same $N_{\rm min}$ regardless of the percentile because the differences in mass between the definitions are relatively modest \citepalias{diemer_20_catalogs}. We now add the $n$ halos from all \LCDM simulations that contribute to a given bin (Table~\ref{table:sims}) and compute the subhalo of flyby fraction. For the self-similar simulations, we follow the same procedure but combine different redshifts at fixed peak height rather than different simulations.

The resolution limit $N_{\rm min}$ depends on which variable we are binning in, on whether we are computing the subhalo or flyby fraction, and on whether we are using \LCDM or self-similar cosmologies. The subhalo fraction demands a higher $N_{\rm min}$ because it relies on the mass or $\vmax$ of subhalos \citep[see also][]{mansfield_21_resolution}. For the flyby fraction, we are counting host halos. Their prior subhalo status depends on larger halos, and the catalogs are complete down to 200 particles for host halos \citepalias{diemer_20_catalogs}. Thus, we set limits of $N_{\rm min} = 500$ and 200 for the subhalo and flyby fractions, respectively. These limits refer to the current bound mass or the equivalent peak height of halos. The halo selection is somewhat trickier with $M_{\rm peak}$ because halos are biased to be hosts near the cut-off of our catalogs at $N_{\rm 200m,peak} \geq 200$. This selection effect occurs due to the different mass evolution of hosts and subhalos: if two halos have the same mass close to the threshold and one becomes a subhalo, that subhalo is more likely to narrowly miss the catalog cut in the future. To restore convergence between simulations with different resolution, we increase $N_{\rm min}$ to $1000$ and $350$ for $\fsub$ and $\ffb$, respectively. When binning in $\vmax$, we encounter yet another selection effect: the $\vmax$ of host halos is tightly correlated with their mass, meaning that our catalog cut in $M_{\rm 200m,peak}$ selects a well-defined range of $\vmax$. For subhalos, however, $\vmax$ does decrease somewhat as they lose mass, leading to the lowest $\vmax$ bins being entirely dominated by subhalos. Again, we find that cuts of $1000$ and $350$ particles are sufficient. Finally, when computing the flyby fraction in self-similar cosmologies, we use $500$ instead of $200$ particles; otherwise, we notice significant non-convergence.

Our choices of $N_{\rm min}$ are informed by studying the convergence between different \LCDM simulations and different redshifts in self-similar simulations; Figure~\ref{fig:conv1} shows representative examples of these comparisons. We do notice residual convergence issues in the flyby fraction in self-similar simulations, but those results are not important for any of our conclusions. 

\subsection{Statistical Uncertainties}
\label{sec:app:err}

In addition to the binned $\fsub$ and $\ffb$, we wish to compute estimates of their statistical uncertainty. In a given bin, we measure the total number of halos, $N_{\rm bin}$, and a subhalo or flyby fraction $f$. The resulting uncertainty is estimated using the binomial formula, $\sigma_{f} = \sqrt{ f (1 - f) / N_{\rm bin}}$. 

We also need to quantify the uncertainty on the ratio between $\fsub$ in some radius definition and in $\rvir$, as shown in the bottom panels of Figure~\ref{fig:frac}. Combining the uncertainties on the respective fractions is a poor estimate because there is a strong correlation between a halo being a subhalo according to two different radius definitions. To take this correlation into account, we need to consider the same halos in each definition. If we want to take this correlation into account, we cannot simply divide the fractions shown in the top panels of Figure~\ref{fig:frac} because they may be based on slightly different sets of halos (due to resolution limits discussed above). Instead, we calculate $\fsub$ based on $\rvir$ for the same halos as for a given other definition, even if some of those halos would be below the rescaled resolution limit for $\rvir$. We then calculate the uncertainty on the ratio by jackknife resampling. 

For the flyby fraction, we compare to $\ffb$ based on $\rfoc$. Performing this calculation in the same way as for $\fsub$ would be complicated due to the criteria for what constitutes a flyby halo (Section~\ref{sec:results:flyby}). Thus, we simply divide the binned $\ffb$ for each radius definition by that for $\rfoc$ and combine the respective binomial errors in quadrature. The resulting uncertainty is an overestimate due to the aforementioned correlation, but we expect the difference to be small because there are generally many fewer flyby halos than when using $\rfoc$. Thus, either the binomial or jackknife uncertainties will be dominated by the uncertainty in the smaller $\ffb$. The size of the error bars has no bearing on our conclusions.


\bibliographystyle{aasjournal}
\bibliography{\includedir/bib_mine.bib,\includedir/bib_general.bib,\includedir/bib_structure.bib,\includedir/bib_galaxies.bib,\includedir/bib_clusters.bib}

\end{document}